\documentclass[sigconf, authorversion]{acmart}

\AtBeginDocument{%
  }

\copyrightyear{2024}
\acmYear{2024}
\setcopyright{acmlicensed}
\acmConference[WWW '24] {Proceedings of the ACM Web Conference 2024}{May 13--17, 2024}{Singapore, Singapore.}
\acmBooktitle{Proceedings of the ACM Web Conference 2024 (WWW '24), May 13--17, 2024, Singapore, Singapore}
\acmISBN{979-8-4007-0171-9/24/05}
\acmDOI{10.1145/3589334.3645486}

\usepackage{enumitem}
\usepackage{subfigure}
\usepackage[linesnumbered,ruled,vlined]{algorithm2e}
\usepackage{amsthm}

\usepackage{amssymb}
\usepackage{amsfonts}
\usepackage{multirow}
\usepackage{float}
\usepackage{bbding}
\usepackage{balance}
\usepackage{hyperref}

\usepackage{color}
\newcommand{\blue}[1]{\textcolor{blue}{#1}}
\definecolor{brown}{RGB}{139,64,0}

\begin{document}

\title{Linear-Time Graph Neural Networks for Scalable Recommendations}

\settopmatter{authorsperrow=4}

\author{Jiahao Zhang}
\authornote{Equal contributions.}
\affiliation{
    \institution{\small{The Hong Kong Polytechnic University}}
    \country{}
}
\email{22037278r@connect.polyu.hk}

\author{Rui Xue}
\authornotemark[1]
\affiliation{
    \institution{\small{North Carolina State University}}
    \country{}
}
\email{rxue@ncsu.edu}

\author{Wenqi Fan}
\authornote{Corresponding author: Wenqi Fan, Department of Computing, and 
 Department of Management and Marketing, The Hong Kong Polytechnic University.}
\affiliation{
    \institution{\small{The Hong Kong Polytechnic University}}
    \country{}
}
\email{wenqi.fan@polyu.edu.hk}

\author{Xin Xu}
\affiliation{
    \institution{\small{The Hong Kong Polytechnic University}}
    \country{}
}
\email{xin.xu@polyu.edu.hk}

\author{Qing Li}
\affiliation{
    \institution{\small{The Hong Kong Polytechnic University}}
    \country{}
}
\email{qing-prof.li@polyu.edu.hk}

\author{Jian Pei}
\affiliation{
    \institution{\small{Duke University}}
    \country{}
}
\email{j.pei@duke.edu}

\author{Xiaorui Liu}
\affiliation{
    \institution{\small{North Carolina State University}}
    \country{}
}
\email{xliu96@ncsu.edu}


%
%



\begin{abstract}
In an era of information explosion, recommender systems are vital tools to deliver personalized recommendations for users. 
The key of recommender systems is to forecast users' future behaviors based on previous user-item interactions. 
Due to their strong expressive power of capturing high-order connectivities in user-item interaction data, recent years have witnessed a rising interest in leveraging Graph Neural Networks (GNNs) to boost the prediction performance of recommender systems. 
Nonetheless, classic Matrix Factorization (MF) and Deep Neural Network (DNN) approaches still play an important role in real-world large-scale recommender systems due to their scalability advantages. 
Despite the existence of GNN-acceleration solutions, it remains an open question whether GNN-based recommender systems can scale as efficiently as classic MF and DNN methods. 
In this paper, we propose a Linear-Time Graph Neural Network (LTGNN) to scale up GNN-based recommender systems to achieve comparable scalability as classic MF  approaches while maintaining GNNs' powerful expressiveness for superior prediction accuracy. 
Extensive experiments and ablation studies are presented to validate the effectiveness and scalability of the proposed algorithm. 
Our implementation based on PyTorch is available~\footnote{\blue{~\url{https://github.com/QwQ2000/TheWebConf24-LTGNN-PyTorch}}}.
\end{abstract}

\maketitle

\section{Introduction}

In an era of information explosion, recommender systems are playing an increasingly critical role in enriching users' experiences with various online applications, due to their remarkable abilities in providing personalized item recommendations. 
The main objective of recommender systems is to predict a list of candidate items that are likely to be clicked or purchased by capturing users' potential interests from their historical behaviors~\citep{he2020lightgcn}. 
A prevailing technique in modern recommender systems is collaborative filtering (CF), which leverages the patterns across similar users and items to predict the users' preferences. 

As one of the most representative CF methods, matrix factorization (MF) models are introduced to represent users and items in a low-dimensional embedding space by encoding the user-item interactions matrix. 
After the emergence of MF models, a remarkable stream of literature has made great efforts to improve the expressive capability of user and item representations. 
As discussed in many previous studies~\citep{wang2019neural, wang2020disentangled, liu2023generative}, we can divide these attempts into two branches based on their modeling ability of user-item interaction graphs. 
First, most early approaches in collaborative filtering focus on the \emph{local connectivity} of users and items, such as item similarity models~\citep{koren2008factorization, he2018nais} and deep neural networks (DNNs)~\citep{he2017neural, xue2017deep}. 
Second, due to the intrinsic limitation of modeling high-order connectivity in early CF models, recent years have witnessed a rising interest in graph neural networks (GNNs) in recommendations. 
To be specific,  GNN-based CF models encode both \emph{local and long-range collaborative signals} into user and item representations by iteratively aggregating embeddings along local neighborhood structures in the interaction graph~\citep{wang2019neural, he2020lightgcn, fan2019graph}, showing their superior performance in modeling complex user-item interaction graphs.

Despite the promising potential of GNNs in modeling high-order information in interaction graphs, GNN-based CF models have not been widely employed in industrial-level applications majorly due to their scalability limitations~\cite{hamilton2017inductive, ying2018graph}. 
In fact, classic CF models like MF and DNNs are still playing major roles in real-world applications due to their computational advantages, especially in large-scale industrial recommender systems~\citep{elkahky2015multi, covington2016deep}. 
In particular, the computation complexity for training these conventional CF models, such as MF and DNNs, is \textit{linear} to the number of user-item interactions in the interaction matrix, while the computation complexity of training GNN-based CF models are \textit{exponential} to the number of propagation layers or \textit{quadratic} to the number of edges (as will be discussed in Section~\ref{sec:gnn}).
    
In web-scale recommender systems, the problem size can easily reach a billion scale towards the numbers of nodes and edges in the interaction graphs~\citep{wang2018billion, jin2023amazon}. 
Consequently, it is essential that scalable algorithms should have nearly linear or sub-linear complexity with respect to the problem size. Otherwise, they are infeasible in practice due to the unaffordable computational cost~\cite{teng2016scalable}. 
While numerous efforts have continued to accelerate the training of GNN-based recommender systems, including two main strategies focusing on neighbor sampling~\citep{ying2018graph, li2020hierarchical} and design simplification~\citep{wu2019simplifying, he2020lightgcn}, none of them can achieve the linear complexity for GNN-based solutions, 
leading to inferior efficiency in comparison with conventional CF methods such as MF and DNNs. 
There is still an open question in academia and industry: \emph{Whether GNN-based recommendation models can scale linearly as the classic MF and DNN methods, while exhibiting long-range modeling abilities and strong prediction performance.}

In this paper, our primary objective revolves around \textit{1) preserving the strong expressive capabilities inherent in GNNs} while simultaneously \textit{2) achieving a linear computation complexity} that is comparable to traditional CF models like MF and DNNs. 
However, it is highly non-trivial to pursue such a scalable GNN design, since the expressive power of high-order collaborative signals lies behind the number of recursive aggregations (i.e., GNN layers). Moreover, the embedding aggregation over a large number of neighbors is highly costly.
To achieve a linear computation complexity, we propose a novel implicit graph modeling for recommendations with the \textit{single-layer propagation} model design and an \textit{efficient variance-reduced neighbor sampling} algorithm. 
Our contributions can be summarized as follows: 
\vspace{-\topsep}
\begin{itemize}[leftmargin=0.2in]
\item We provide a critical complexity analysis and comparison of widely used collaboration filtering approaches, and we reveal their performance and efficiency bottlenecks.

\item We propose a novel GNN-based model for large-scale collaborative filtering in recommendations, namely LTGNN (Linear Time Graph Neural Networks), which only incorporates \textit{one propagation layer} while preserving the capability of capturing long-range collaborative signals. 
    
\item To handle large-scale user-item interaction graphs, we design an efficient and improved \textit{variance-reduced neighbor sampling} strategy for LTGNN to significantly reduce the neighbor size in embedding aggregations. The random error caused by neighbor sampling is efficiently tackled by our improved variance reduction technique. 
    
\item We conduct extensive comparison experiments and ablation studies on three real-world recommendation datasets, including a large-scale dataset with millions of users and items. The experiment results demonstrate that our proposed LTGNN significantly reduces the training time of GNN-based CF models while preserving the recommendation performance on par with previous GNN models. We also perform detailed time complexity analyses to show our superior efficiency.
\end{itemize}
\vspace{-\topsep}
\section{Preliminaries}
This section presents the notations used in this paper, and then briefly introduces preliminaries about GNN-based recommendations and the computation complexity of popular CF models.

\subsection{Notations and Definitions}

In personalized recommendations, the historical user-item interactions can be naturally represented as a bipartite graph $\mathcal{G}=(\mathcal{V}, \mathcal{E})$, where the node set $\mathcal{V}$ includes $n$ user nodes $\{v_1, \cdots, v_n\}$ and $m$ item nodes $\{v_{n+1}, \cdots, v_{n+m}\}$, and the edge set $\mathcal{E}=\{e_1, \cdots, e_{|\mathcal{E}|}\}$ consists of undirected edges between user nodes and item nodes. 
It is clear that the number of undirected edges $|\mathcal{E}|$ equals to the number of observed user-item interactions $|\mathcal{R}^+|$ in the training data (i.e., $|\mathcal{E}|=|\mathcal{R}^+|$). 
The graph structure of $\mathcal{G}$ can be denoted as the adjacency matrix $\boldsymbol{A}\in\mathbb{R}^{(n+m)\times (n+m)}$, and its diagonal degree matrix are denoted as $\boldsymbol{D}$. 
The normalized adjacency matrix with self-loops is defined as $\boldsymbol{\tilde{A}}=(\boldsymbol{D}+\boldsymbol{I})^{-\frac{1}{2}}(\boldsymbol{A}+\boldsymbol{I})(\boldsymbol{D}+\boldsymbol{I})^{-\frac{1}{2}}$.
We use $\mathcal{N}(v)$ to denote the set of neighboring nodes of a node $v$, including $v$ itself.
In addition, the trainable embeddings of user and item nodes in graph $\mathcal{G}$ are denoted as $\boldsymbol{E}=[\boldsymbol{e}_1, \dots,\boldsymbol{e}_n,\ \ \boldsymbol{e}_{n+1}, \dots,\boldsymbol{e}_{n+m}]^T\in\mathbb{R}^{(n+m)\times d}$, where its first $n$ rows are $d$-dimensional user embeddings and its $n+1$ to $n+m$ rows are $d$-dimensional item embeddings.

In the training process of GNN-based collaborative filtering models, we use $(\boldsymbol{E}_l^k)_{\boldsymbol{B}}$ or $(\boldsymbol{e}_l^k)_{v}$ to denote an embedding matrix or a single embedding vector, 
where $k$ is the index of training iterations and $l$ is the index of propagation layers. 
The subscript $(\cdot)_{\boldsymbol{B}}$ or $(\cdot)_v$ denotes the embedding for a batch of nodes $\boldsymbol{B}$ or a single node $v$. 

\subsection{Mini-batch Training}
\label{sec:batch_training}
To provide effective item recommendations from user-item interactions, a typical training objective is the pairwise loss function.
We take the most widely adopted BPR~\cite{rendle2009bpr} loss as an example: 
\begin{equation}\label{eq:bpr}
    \mathcal{L}_{BPR} = \sum_{(u, i, j)\in\mathcal{O}} -\ln \sigma(\hat{y}_{u, i} - \hat{y}_{u, j}), 
\end{equation}
where $\mathcal{O}=\{(u, i, j) | (u, i)\in\mathcal{R}^{+}, (u, j)\in\mathcal{R}^{-}\}$ denotes the pairwise training data. 
$\mathcal{R}^{+}$ and $\mathcal{R}^{-}$ denotes observed and unobserved user-item interactions. 
In practice, the training data $\mathcal{O}$ is hardly leveraged in a full-batch setting due to the large number of user-item interations~\cite{he2020lightgcn, wang2019neural}. 
Therefore, mini-batch training is a common choice that splits the original data $\mathcal{O}$ into multiple components $\boldsymbol{\Omega} = \{\mathcal{O}_{(u_1, i_1)}, \mathcal{O}_{(u_2, i_2)}, \cdots, \mathcal{O}_{(u_{|\mathcal{R}^{+}|}, i_{|\mathcal{R}^{+}|})}\}$, where $\mathcal{O}_{(u_r, i_r)}=\{(u_r, i_r, j) | (u_r, j)\in\mathcal{R}^-\}$ contains all the training data including positive and negative samples for a specific interaction $(u_r,i_r)$. 
In each training iteration, we first sample $B$ interactions from $\mathcal{R}^+$, which is denoted as $\hat{\mathcal{R}^+}$, satisfying $|\hat{\mathcal{R}^+}|=B$. 
Afterward, we create the training data for $\hat{\mathcal{R}^+}$ by merging the corresponding components in $\boldsymbol{\Omega}$, which can be denoted as $\hat{\mathcal{O}}(\hat{\mathcal{R}}^+) = \bigcup_{(u, i)\in\hat{\mathcal{R}}^+} \mathcal{O}_{(u,i)}$. 
Thus, the mini-batch training objective can be formalized as follows:
\begin{equation}\label{eq:bpr_batch}
    \hat{\mathcal{L}}_{BPR}(\hat{\mathcal{R}^+}) = \sum_{(u, i, j)\in\hat{\mathcal{O}}(\hat{\mathcal{R}}^+)} -\ln \sigma(\hat{y}_{u, i} - \hat{y}_{u, j}). 
\end{equation}
In each training epoch, we iterate over all user-item interactions in $\mathcal{R}^+$, so the mini-batch training objective $\hat{\mathcal{L}}_{BPR}$ needs to be evaluated for $|\mathcal{R}^+| / B$ times (i.e., $|\mathcal{E}| / B$ times). 

\subsection{GNNs and MF for Recommendations} 
\label{sec:gnn}

In this subsection, we will briefly review MF and two representative GNN-based recommendation models, including LightGCN~\cite{he2020lightgcn} and PinSAGE~\cite{ying2018graph}, and discuss their computation complexity. 

\noindent  {\textbf{LightGCN}.} 
Inspired by the graph convolution operator in GCN~\cite{kipf2016semi} and SGC~\cite{wu2019simplifying}, LightGCN~\cite{he2020lightgcn} iteratively propagates the user embedding $(\boldsymbol{e}_l)_u$ and item embedding $(\boldsymbol{e}_l)_i$ as follows:
\begin{align}\label{eq:lgn_user}
    (\boldsymbol{e}_{l+1})_u = \frac{1}{\sqrt{|\mathcal{N}(u)|}}\sum_{i\in \mathcal{N}(u)}\frac{1}{\sqrt{|\mathcal{N}(i)|}}(\boldsymbol{e}_l)_i, \\
    \label{eq:lgn_item}
     (\boldsymbol{e}_{l+1})_i = \frac{1}{\sqrt{|\mathcal{N}(i)|}}\sum_{i\in \mathcal{N}(i)}\frac{1}{\sqrt{|\mathcal{N}(u)|}}(\boldsymbol{e}_l)_u.
\end{align}
The embedding propagation of LightGCN can be re-written in matrix form as follows:
\begin{align}
\label{eq:sgc_prop}    \boldsymbol{E}_{l+1} &=\boldsymbol{\tilde{A}}\boldsymbol{E}_l, \quad\forall l=0,\dots, L - 1  \\
\boldsymbol{Y} &= \frac{1}{L + 1}\sum_{l=0}^L\boldsymbol{E}_l, 
\end{align}
where $L$ denotes the number of GNN layers, and $\boldsymbol{Y}$ denotes the model output of LightGCN with layer-wise combination. 
As LightGCN computes full embedding propagation in Eq.~\eqref{eq:sgc_prop} for $L$ times to capture $L$-hop neighborhood information, the computation complexity of LightGCN in one training iteration is $\mathcal{O}(L|\mathcal{E}|d)$ with the support of sparse matrix multiplications. Thus, the computation complexity for one training epoch is $\mathcal{O}(\frac{|\mathcal{E}|}{B}\cdot L|\mathcal{E}|d) = \mathcal{O}(\frac{1}{B}L|\mathcal{E}|^2d)$, where $\frac{|\mathcal{E}|}{B}$ is the number of training iterations in one epoch.

\noindent  {\textbf{PinSAGE}.}
The embedding propagation in LightGCN aggregates all the neighbors for a user or an item, which is less compatible with Web-scale item-to-item recommender systems. 
Another important embedding propagation rule in GNN-based recommendation is proposed in PinSAGE:
\begin{align}\label{eq:sage_prop}
    (\boldsymbol{n}_{l+1})_u &= \text{Aggregate}(\{\text{ReLU}(\boldsymbol{Q}\cdot(\boldsymbol{e}_{l})_v+\boldsymbol{q})\ |\ v\in\hat{\mathcal{N}}(u)\}), \\
    (\boldsymbol{e}_{l+1})_u &= \text{Normalize}(\boldsymbol{W}\cdot\text{Concat}[(\boldsymbol{e}_{l})_u; (\boldsymbol{n}_{l+1})_u]+\boldsymbol{w}),
\end{align}
where $\boldsymbol{Q}, \boldsymbol{q}, \boldsymbol{W}, \boldsymbol{w}$ are trainable parameters, and $\hat{\mathcal{N}}(u)$ denotes the randomly sampled neighbors for node $u$. 
If PinSAGE constantly samples $D$ random neighbors for each node at each layer, and the sampled $B$ edges have $n_B$ target nodes without repetition, 
the computation complexity in each training iteration is $\mathcal{O}(n_BD^Ld^2)$ as discussed in previous studies~\cite{wu2020comprehensive}. 
Thus, the time complexity in the entire training epoch is $\mathcal{O}(\frac{|\mathcal{E}|}{B}\cdot n_BD^Ld^2)$ = $\mathcal{O}(|\mathcal{E}|D^Ld^2)$, as $n_B$ and $B$ shares the same order. Moreover, the neighbor sampling in PinSAGE incurs large approximation errors that impact the prediction performance.

\noindent  {\textbf{Matrix Factorization (MF)}.} 
Matrix factorization and its neural variant NCF~\cite{he2017neural} are simple but strong baselines for recommendations at scale. 
Given learnable user embedding $\boldsymbol{p}_u$ and item embedding $\boldsymbol{q}_i$, MF models their interaction directly by inner product as $\hat{y}_{u, i} = \boldsymbol{p}_u^T\boldsymbol{q}_i$, while NCF models the interaction by deep neural networks as follows:
\begin{align}\label{eq:ncf1}
    \boldsymbol{e}_L &= \phi(\boldsymbol{W}_L(\cdots\phi(\boldsymbol{W}_2\phi(\boldsymbol{W}_1\begin{bmatrix}\boldsymbol{p}_u\\ \boldsymbol{q}_i\end{bmatrix} + \boldsymbol{b_1})+\boldsymbol{b}_2)\cdots)+\boldsymbol{b}_L), \\
    \label{eq:ncf1}
    \hat{y}_{u, i} &= \sigma(\boldsymbol{h}^T\boldsymbol{e}_L),
\end{align}
where $\boldsymbol{W}$, $\boldsymbol{b}$ and $\boldsymbol{h}$ are trainable parameters, and $\phi$ is a non-linear activation function. In each training iteration, the computation complexity for MF and NCF is $\mathcal{O}(Bd)$ and $\mathcal{O}(BLd^2)$, which stands for the complexity of dot products and MLPs, respectively. 
Thus, the time complexity in each training epoch for MF and NCF is $\mathcal{O}(|\mathcal{E}|d)$ and $\mathcal{O}(|\mathcal{E}|Ld^2)$.

\noindent {\textbf{Inefficiency of GNNs.}} 
In comparison with conventional MF models, GNNs' inefficiency lies behind their non-linear complexity with respect to the number of edges $|\mathcal{E}|$ or layers $L$. 
For example, the time complexity for LightGCN is $\mathcal{O}(\frac{1}{B}L|\mathcal{E}|^2d)$, which grows quadratically with $|\mathcal{E}|$, and PinSAGE has a complexity of $\mathcal{O}(|\mathcal{E}|D^Ld^2)$, which grows exponentially with $L$. 
In this paper, we pursue a \emph{linear-time} design for GNNs, which means the time complexity of our proposed model is expected to be $\mathcal{O}(C|\mathcal{E}|d)$, where $C$ is a small constant (e.g., $C=Ld$ for NCF).

\begin{figure*}[t]
\vspace{-0.38in}
    \centering
    \includegraphics[width=0.9\linewidth]{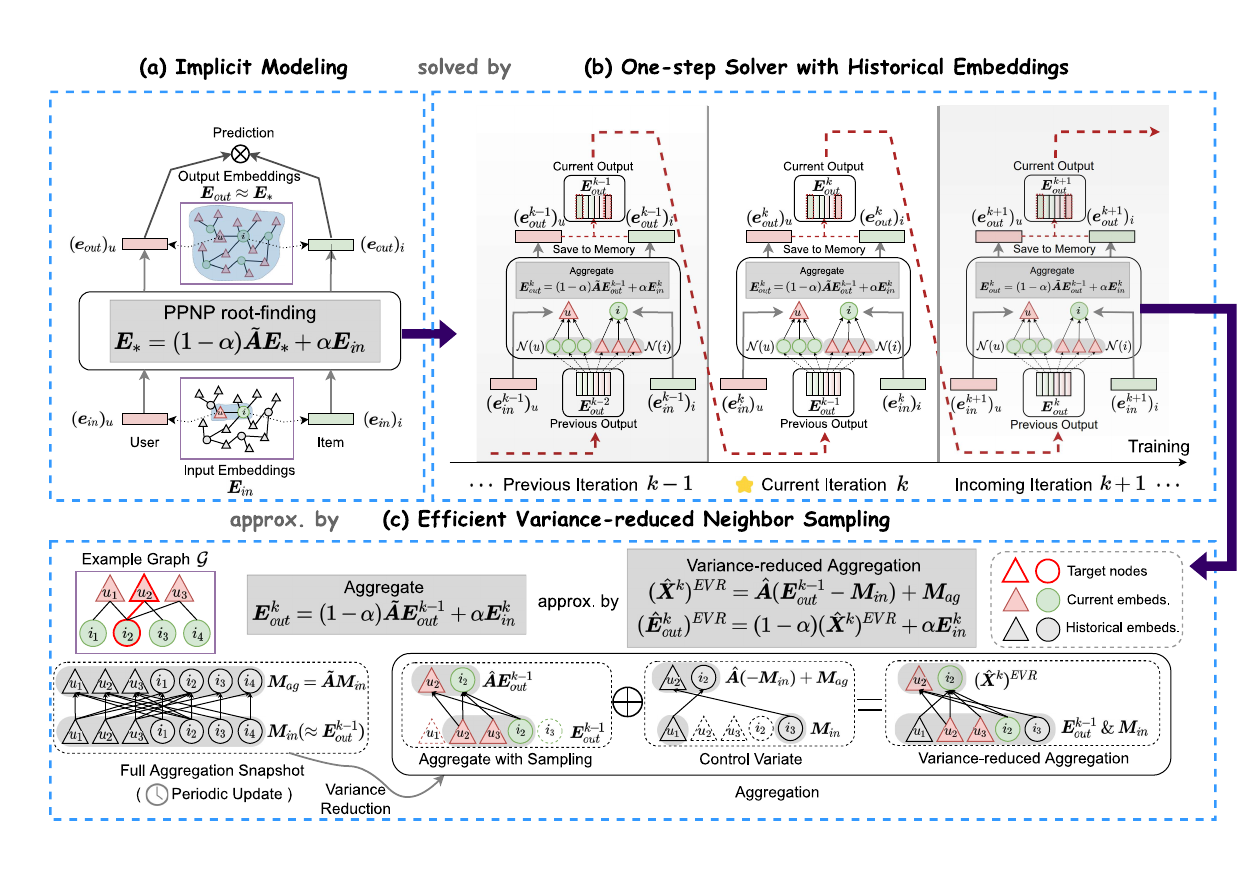}
    \vspace{-0.4in}
    \caption{An illustration of our model architecture. (a) The forward process of our model aims to solve the PPNP fixed-point equation, which expresses an equilibrium state of the embedding propagations, and can be used to capture long-range relations between any pair of nodes regardless of their distance. (b) The PPNP fixed-point equation is solved with a single forward propagation layer, which leverages the historical output embeddings in previous training iterations. (c) The process of efficient variance-reduced neighbor sampling in LTGNN.}
    \label{fig:banner}
\vspace{-0.2in}
\end{figure*}

\newcommand{\vE}{\boldsymbol{E}}
\newcommand{\vI}{\boldsymbol{I}}
\newcommand{\vA}{\boldsymbol{A}}
\newcommand{\tA}{\tilde{\boldsymbol{A}}}
\newcommand{\cL}{\mathcal{L}}
\newcommand{\cO}{\mathcal{O}}
\newcommand{\cE}{\mathcal{E}}

\section{The Proposed Method}\label{sec:proposed}
The scalability issue of GNN-based recommendation models inspires us to pursue a more efficient algorithm design with linear computation complexities. 
However, to reduce the computation complexity while preserving GNNs' long-range modeling ability, we need to overcome \textbf{two main challenges}: 
\vspace{-\topsep}
\begin{itemize}
[leftmargin=0.2in]
\item \textbf{Layer and expressiveness}: Increasing the number of layers $L$ in GNNs can capture long-range dependencies, but the complexity (e.g., $\mathcal{O}(|\mathcal{E}|D^Ld^2)$ in PinSAGE) can hardly be linear when $L$ is large. 
\item \textbf{Neighbor aggregation and random error}: The number of neighbors $D$ aggregated for each target node in a GNN layer substantially affects both computation cost and approximation error. Aggregating all neighbors (e.g., LightGCN) is costly, while aggregation with random sampling incurs large errors (e.g., PinSAGE).
\end{itemize}
\vspace{-\topsep}
\noindent To address these challenges, we propose implicit graph modeling in Section~\ref{sec:impl_modeling} to reduce the layer number $L$ significantly, and a variance-reduced neighbor sampling strategy in Section~\ref{sec:vr} to lower the neighbor aggregation cost for high-degree nodes.
We carefully handle the numerical and random errors of our designs, and ensure the strong expressiveness of our proposed LTGNN.

\subsection{Implicit Modeling for Recommendations}\label{sec:impl_modeling} 

Personalized PageRank~\cite{lawrence1998pagerank} is a classic approach for the measurement of the proximity between nodes in a graph. 
It is adopted by a popular GNN model, PPNP (Personalized Propagation of Neural Predictions)~\cite{gasteiger2018predict}, to propagate node embeddings according to the personalized PageRank matrix:
\begin{align}
\label{eq:ppnp}
\vE_{PPNP}^k = \alpha \Big ( \vI - (1-\alpha) \tA \Big)^{-1} \vE_{in}^k,
\end{align}
where $\alpha$ is the teleport factor and $\vE_{in}^k$ is the input node embedding.
Due to the infeasible cost of matrix inversion, APPNP approximates this by $L$ propagation layers:
\begin{align}
\label{eq:mp_in}
\vE_{0}^k &= \vE^{k}_{in},\\
\vE_{l+1}^k &= (1-\alpha) \tA \vE_{l}^k+\alpha \vE_{in}^k, ~\forall l=0,\dots,L-1  \label{eq:mp_iter}
\end{align}
such that it can capture the $L$-hop high-order information in the graph without suffering from over-smoothing due to the teleport term $\alpha \vE_{in}^k$. 
LightGCN exhibits a close relation with APPNP although the embedding from different propagation layers is averaged with a different weight (see the analysis in Section 3.2 of~\cite{he2020lightgcn}). 
However, like most GNNs, both APPNP and LightGCN suffer from scalability issues due to the multi-layer recursive feature aggregations, which greatly limit their applications in large-scale recommendations.  

Motivated by the previous success of Implicit Deep Learning~\cite{el2021implicit} and Implicit GNNs~\cite{gu2020implicit, li2022unbiased, xue2023lazygnn}, 
we propose implicit modeling for graph-based recommendations (as shown in Fig.~\ref{fig:banner}(a)), which directly computes the fixed point $\vE_*^k$ of the embedding propagation in Eq.~\eqref{eq:mp_iter}. 
Particularly, the output of our implicit model can be formalized as the solution of a linear system: 
\begin{align}
\label{eq:implicit-fix-point}
\vE^k_{out} = \text{RootFind}(\vE^k_*),\quad \text{s.t.}\quad \vE^k_{*}=(1-\alpha)\tilde{\vA}\vE^k_{*}+\alpha\vE_{in}^k,  
\end{align}
where the relation between output embedding $\vE_{out}^k$ and input embedding $\vE_{in}^k$ is implicitly defined by a root-finding process of the fixed-point equation. 
Formulating graph-based recommendations implicitly with a fixed-point equation has two advantages: 1) The fixed-point $\vE^k_*$ models the equilibrium state of embedding propagations, which is equivalent to the extreme state under an infinite number of propagations, effectively capturing the long-range node dependencies in graphs. 
2) This implicit modeling provides flexibility for the GNN design, as we can use any equation solver to acquire $\vE^k_*$ instead of stacking multiple GNN layers. 

Specifically, to pave the way to linear-time computation, we propose to solve this fixed-point equation by a single forward propagation layer, as shown in Fig.~\ref{fig:banner}(b):
\begin{align}
\vE_{out}^k &= (1-\alpha) \tA \vE_{out}^{k-1}+\alpha \vE_{in}^k,
\label{eq:1layer_solver}
\end{align}
where $\vE_{out}^{k-1}$ is the historical output embeddings at previous training iteration $k-1$ and serves as a better initialization for the fixed-point solver. 
This single-layer design is ultra-efficient compared with multi-layer embedding propagations but still captures multi-hop neighbor information through information accumulation across training iterations.

The backward propagation of implicit models is independent of the forward computation~\cite{el2021implicit, gu2020implicit, xue2023lazygnn}. Given the gradient from the output embedding layer $\frac{\partial \cL}{\partial \vE_{out}^k}$, the gradient of $\vE_{in}^k$ can be computed based on the fixed-point equation in Eq.~\eqref{eq:implicit-fix-point}:
\begin{align}
\frac{\partial \cL}{\partial \vE_{in}^k} = \alpha \frac{\partial \cL}{\partial \vE_{out}^k}
\Big ( \vI - (1-\alpha) \tA \Big)^{-1}.
\end{align}

\noindent Due to the prohibitively high dimensionality of the adjacency matrix, computing its inverse is infeasible. Therefore, we propose to approximate this gradient by a single backward propagation layer:
\begin{align} 
\frac{\partial \cL}{\partial \vE_{in}^k}  &= (1-\alpha) \tA \frac{\partial \cL}{\partial \vE_{in}^{k-1}}  + \alpha \frac{\partial \cL}{\partial \vE_{out}^k},
\end{align}
where $\frac{\partial \cL}{\partial \vE_{in}^{k-1}}$ is the historical gradient of input embedding from iteration $k-1$ and serves as a better initialization for the fixed-point solver. 
In summary, the forward and backward computation of our single-layer GNN are formulated as:
\begin{align}
\label{eq:forward_backward1}
\text{\textbf{Forward}:}~~~ \vE_{out}^k &= (1-\alpha) \tA \vE_{out}^{k-1}+\alpha \vE_{in}^k, \\
\text{\textbf{Backward}:} ~~~ \frac{\partial \cL}{\partial \vE_{in}^k}  &= (1-\alpha) \tA \frac{\partial \cL}{\partial \vE_{in}^{k-1}}  + \alpha \frac{\partial \cL}{\partial \vE_{out}^k},
\label{eq:forward_backward}
\end{align}
where the historical computations $\vE_{out}^{k-1}$ and $\frac{\partial \cL}{\partial \vE_{in}^{k-1}}$ can be obtained by maintaining the model outputs and gradients at the end of each training iteration.

\subsection{Efficient Variance-Reduced Neighbor Sampling}
\label{sec:vr}

The previous section presents an implicit modeling and single-layer design for GNNs, which significantly reduces GNNs' computation complexity. For example, PinSAGE has a complexity of $\cO(|\cE|Dd^2)$ given there is only one layer ($L=1$), which can be linear if $D$ is a small constant. Unfortunately, $D$ affects the expressiveness of GNNs in recommendations and cannot be easily lowered.

As explained in Section~\ref{sec:batch_training}, the mini-batch training process samples user-item interactions (i.e., links in the user-item graph) in each iteration, which means that nodes with higher degrees are more likely to be sampled. These nodes also need more neighbors to compute their output embeddings accurately. Therefore, a small $D$ will introduce large approximation errors and degrade the performance. This is consistent with previous studies on the impact of neighbor sampling in large-scale OGB benchmarks~\cite{duan2022comprehensive}. Some methods, such as VR-GCN~\cite{chen2018stochastic} and MVS-GNN~\cite{cong2020minimal}, use variance-reduction (VR) techniques to reduce the random error in neighbor sampling. However, we will show that these methods still require the full aggregation of historical embeddings, which maintains an undesirable complexity. To address this issue, we will propose an efficient VR neighbor sampling approach that achieves linear complexity while controlling the random error.

\noindent  \textbf{Classic Variance-reduced Neighbor Aggregation.} 
Recent research has investigated variance reduction on GNNs, such as VR-GCN and MVS-GNN~\cite{chen2018stochastic,cong2020minimal}:
\begin{align}\label{eq:vector_vr_prop1}
    (\hat{\boldsymbol{X}}^k)^{VR}&= \hat{\boldsymbol{A}}(\boldsymbol{E}_{in}^k - \boldsymbol{\overline{E}}^k_{in}) + \tilde{\boldsymbol{A}}\boldsymbol{\overline{E}}^k_{in} \quad (\approx \tilde{\boldsymbol{A}}\boldsymbol{E}^k_{in}), 
\end{align}

\noindent where $\hat{\boldsymbol{A}}$ is an unbiased estimator of $\tilde{\boldsymbol{A}}$, $\hat{\boldsymbol{A}}_{u,v} = \frac{|\mathcal{N}(u)|}{D} \tilde{\boldsymbol{A}}_{u,v}$ if node $v$ is sampled as a neighbor of target node $u$, otherwise $\hat{\boldsymbol{A}}_{u,v} = 0$. $\overline{\boldsymbol{E}}^k_{in}$ is the historical embeddings for approximating $\boldsymbol{E}^k_{in}$. However, such approaches need to perform full neighbor aggregations on the historical embedding by computing $\tilde{\boldsymbol{A}}(\boldsymbol{\overline{E}}^k_{in})$.
Importantly, this computation has to be performed in each mini-batch iteration, leading to the quadratic computation complexity $\cO(\frac{|\cE|^2d}{B})$ for the whole training epoch. 
Therefore, they seriously sacrifice the computational efficiency of neighbor sampling in large-scale recommender systems. 
Besides, it is noteworthy that other GNN acceleration approaches based on historical embeddings~\cite{fey2021gnnautoscale, yu2022graphfm} also suffer from the full aggregation problem.

\noindent \textbf{Efficient Variance-reduced Neighbor Sampling.} 
To further reduce the quadratic computation complexity, we propose to compute the historical embedding aggregation periodically instead of computing them in every training iteration. Specifically, we allocate two memory variables $\boldsymbol{M}_{in}$ and $\boldsymbol{M}_{ag}$ to store the historical input embeddings and fully aggregated embeddings, 
where $\boldsymbol{M}_{ag} = \tilde{\boldsymbol{A}}\boldsymbol{M}_{in}$. 
The input memory variable $\boldsymbol{M}_{in}$ is updated periodically at the end of each training epoch, and the aggregated embedding $\boldsymbol{M}_{ag}$ are updated based on the renewed inputs. We name it as Efficient Variance Reduction (EVR), which can be formulated as:
\begin{align} \label{eq:mat_vr_prop1}
(\hat{\boldsymbol{X}}^k)^{EVR} &= \hat{\boldsymbol{A}}(\boldsymbol{E}^{k-1}_{out} - \boldsymbol{M}_{in}) + \boldsymbol{M}_{ag} &(\approx \tilde{\boldsymbol{A}}\boldsymbol{E}^{k-1}_{out}),\\
    \label{eq:mat_vr_prop2}(\hat{\boldsymbol{E}}^k_{out})^{EVR} &= (1-\alpha) (\hat{\boldsymbol{X}}^k)^{EVR} + \alpha \boldsymbol{E}^k_{in} &(\approx\boldsymbol{E}^k_{out}),
\end{align}
where the first term in Eq.~\eqref{eq:forward_backward1} is approximated by $(\hat{\boldsymbol{X}}^k)^{EVR}$, and the second term remains unchanged. 
An illustration of this sampling algorithm is shown in Fig.~\ref{fig:banner}(c). 

This variance reduction method can also be adapted to backward computations. 
Symmetrically, the backward computation in Eq.~\eqref{eq:forward_backward} can be computed with our proposed EVR as:
\begin{align}
\label{eq:back_vr_prop1}
(\hat{\boldsymbol{G}}^k)^{EVR} &= \hat{\boldsymbol{A}}(\frac{\partial\cL}{\partial\boldsymbol{E}^{k-1}_{in}} - \boldsymbol{M}_{in}') + \boldsymbol{M}_{ag}' &(\approx \tilde{\boldsymbol{A}}\frac{\partial\cL}{\partial\boldsymbol{E}^{k-1}_{in}}),\\
(\frac{\widehat{\partial \cL}}{\partial \boldsymbol{E}_{in}^k})^{EVR}  &= (1-\alpha)(\hat{\boldsymbol{G}}^k)^{EVR}  + \alpha \frac{\partial \cL}{\partial \boldsymbol{E}_{out}^k} &(\approx \frac{\partial\cL}{\partial\boldsymbol{E}^{k}_{in}}), 
\label{eq:back_vr_prop2}
\end{align}
where $\boldsymbol{M}_{in}'$ stores the historical input gradients and $\boldsymbol{M}_{ag}'=\tilde{\boldsymbol{A}}\boldsymbol{M}_{in}'$ maintains the fully aggregated gradients. 
Extra implementation details of LTGNN can be found in Appendix.~\ref{sec:append_model}.

\noindent \textbf{Complexity analysis.} In each training epoch, the efficiency bottleneck lies in the forward and backward computations, which costs $\mathcal{O}(n_BDd)$, as we compute the variance-reduced neighbor aggregation for $n_B$ target nodes, where each target node has $D$ random neighbors. 
Thus, the overall complexity of our method is $\cO(\frac{|\cE|}{B}\cdot n_BDd) = \cO(|\cE|Dd)$, as each training epoch includes $\frac{|\cE|}{B}$ iterations. 
Given that $D$ is a small constant, the complexity no longer preserves an undesirable dependence on $|\cE|^2$ or $L$, and instead, it becomes linear. As a result, it significantly reduces the computational cost in comparison to previous GNN-based recommendation models, as shown in Table.~\ref{tab:complexity}. 

\begin{table}[th]
\vspace{-0.13in}
\caption{Complexity Comparisons.}
\vspace{-0.17in}
\label{tab:complexity}
\begin{center}
\scalebox{0.95}{
\begin{tabular}{c|c}
\hline
\textbf{Models} & \textbf{Computation Complexity} \\ \hline
\textbf{LightGCN}    & $\mathcal{O}(\frac{1}{B}L|\mathcal{E}|^2d)$               \\
\textbf{PinSAGE}      & $\mathcal{O}(|\mathcal{E}|D^Ld^2)$                \\ 
\textbf{MF}      &      $\mathcal{O}(|\mathcal{E}|d)$             \\ 
\textbf{NCF}      &      $\mathcal{O}(L|\mathcal{E}|d^2)$             \\
\textbf{LTGNN}      & $\cO(|\cE|Dd)$  \\ 

\hline
\end{tabular}
}
\end{center}
\vspace{-0.3in}
\end{table}

\section{Experiments}

\begin{table*}[!htb]
\vskip -0.1in
\caption{The comparison of overall prediction performance.}
\vskip -0.15in
\label{tab:comparsion_perf}
\begin{center}
\scalebox{0.95}
{
\begin{tabular}{l|c c|c c|c c}
\hline
\textbf{Dataset}  & \multicolumn{2}{c|}{\textbf{Yelp2018}} & \multicolumn{2}{c|}{\textbf{Alibaba-iFashion}} & \multicolumn{2}{c}{\textbf{Amazon-Large}} \\ \hline
\textbf{Method}  & \textbf{Recall@20} & \textbf{NDCG@20} & \textbf{Recall@20} & \textbf{NDCG@20} & \textbf{Recall@20} & \textbf{NDCG@20} \\ \hline\hline
MF & 0.0436 & 0.0353 & 0.05784 & 0.02676 & 0.02752 & 0.01534\\ 
NCF & 0.0450 & 0.0364 & 0.06027 & 0.02810 & 0.02785 & 0.01807\\ \hline
GC-MC & 0.0462 & 0.0379 & 0.07738 & 0.03395 & OOM & OOM\\ 
PinSAGE & 0.04951 & 0.04049 & 0.07053 & 0.03186 & 0.02809 & 0.01973\\ 
NGCF & 0.0581 & 0.0475 & 0.07979 & 0.03570 & OOM & OOM\\ 
DGCF & 0.064 & 0.0522 & 0.08445 & 0.03967 & OOM & OOM
 \\ 
{LightGCN (L=3)} & \underline{0.06347} & \underline{0.05238} & {0.08793} &	{0.04096} & \textbf{0.0331} & \underline{0.02283}
 \\ \hline 
LightGCN-NS (L=3) & 0.06256 & 0.05140 & \underline{0.08804} & \underline{0.04147} & 0.02835 & 0.02035\\ 
LightGCN-VR (L=3) & 0.06245 & 0.05141 & 0.08814 &	0.04082 & 0.02903 & 0.02093\\ 
LightGCN-GAS (L=3) & 0.06337 & 0.05207 & 0.08169 &	0.03869 & 0.02886 & 0.02085
 \\ \hline\hline
{LTGNN (L=1)} & \textbf{0.06393} & \textbf{0.05245} & \textbf{0.09335} & \textbf{0.04387} & \underline{0.02942} & \textbf{0.02585}\\ \hline
\end{tabular}
}
\end{center}
\vskip -0.15in
\end{table*}

In this section, we will verify the effectiveness and efficiency of the proposed LTGNN framework with comprehensive experiments. 
Specifically, we aim to answer the following research questions:
\vspace{-\topsep}
\begin{itemize}
[leftmargin=0.2in]
    \item \textbf{RQ1}: Can LTGNN achieve promising prediction performance on large-scale recommendation datasets? (Section \ref{sec:perf}) 
    \item \textbf{RQ2}: Can LGTNN handle large user-item interaction graphs more efficiently than existing GNN approaches?  (Section \ref{sec:scale}) 
    \item \textbf{RQ3}: How does the effectiveness of the proposed LTGNN vary when we ablate different parts of the design? (Section \ref{sec:abl}) 
\end{itemize}
\vspace{-\topsep}

\subsection{Experimental Settings}\label{sec:exp_setting}
We first introduce the datasets, baselines, evaluation metrics, and hyperparameter settings as follows. 

\begin{table}[thbp]
\vskip -0.15in
\caption{Dataset statistics.}
\vskip -0.2in
\label{tab:dataset}
\begin{center}
\scalebox{0.95}{
\begin{tabular}{c|cccc}
\hline
\textbf{Dataset}     & \textbf{\# Users} & \textbf{\# Items} & \textbf{\# Interactions} \\ \hline
\textbf{Yelp2018}    & 31, 668   &   38, 048   &   1, 561, 406             \\
\textbf{Alibaba-iFashion}      & 300, 000  &  81, 614  &  1, 607, 813                       \\ 
\textbf{Amazon-Large}   & 872, 557            & 453, 553        & 15, 236, 325                      \\ 
\hline
\end{tabular}
}
\end{center}
\vskip -0.17in
\end{table}

\noindent \textbf{Datasets.}
We evaluate the proposed LTGNN and baselines on two medium-scale datasets, including \textit{Yelp2018} and \textit{Alibaba-iFashion}, and one large-scale dataset \textit{Amazon-Large}. 
Yelp2018 dataset is released by the baselines NGCF~\citep{wang2019neural} and LightGCN~\citep{he2020lightgcn}, and the Alibaba-iFashion dataset can be found in the GitHub repository\footnote{\url{https://github.com/wenyuer/POG}}. 
For the large-scale setting, we construct the large-scale dataset, Amazon-Large, based on the rating files from the Amazon Review Data website\footnote{\url{https://cseweb.ucsd.edu/~jmcauley/datasets/amazon_v2/}}. 
Specifically, we select the three largest subsets (i.e., Books, Clothing Shoes and Jewelry, Home and Kitchen) from the entire Amazon dataset, and then keep the interactions from users who are shared by all the three subsets (7.9\% of all the users). 
The rating scale is from 1 to 5, and we transform the explicit ratings into implicit interactions by only keeping the interactions with ratings bigger than 4.
To ensure the quality of our Amazon-Large dataset, we follow a widely used 10-core setting~\cite{he2017neural,wang2019neural, wang2019kgat} and remove the users and items with interactions less than 10. 
The statistical summary of the datasets can be found in Table~\ref{tab:dataset}.

\noindent  \textbf{Baselines.} 
The main focus of this paper is to enhance the scalability of GNN-based collaborative filtering methods. 
Therefore, we compare our method with the most widely used GNN backbone in recommendations, \textit{LightGCN}~\cite{he2020lightgcn} and its scalable variants that employ typical GNN scalability techniques, including GraphSAGE~\cite{hamilton2017inductive}, VR-GCN~\cite{chen2018stochastic} and GAS~\cite{fey2021gnnautoscale}. 
The corresponding variants of LightGCN are denoted as \textit{LightGCN-NS} (for \underline{N}eighbor \underline{S}ampling), \textit{LightGCN-VR}, and \textit{LightGCN-GAS}.

To demonstrate the effectiveness of our method, we also compare it with various representative recommendation models, including MF~\citep{koren2008factorization}, NCF~\citep{he2017neural}, GC-MC~\citep{berg2017graph}, PinSAGE~\citep{ying2018graph}, NGCF~\citep{wang2019neural}, and DGCF~\citep{he2020lightgcn}.
Moreover, since we are designing an efficient collaborative filtering backbone that is independent of the loss function, our method is orthogonal to SSL-based methods~\cite{wu2021self, yu2022graph} and negative sampling algorithms~\cite{mao2021simplex, huang2021mixgcf}. 
We will explore the combination of our method and these orthogonal designs in future work. Extra comparison results with one of the latest accuracy-driven GNN backbones can be found in Appendix.~\ref{sec:append_exp}.

\noindent  \textbf{Evaluation and Parameter Settings.} Due to the limited space, more implementation details are presented in Appendix.~\ref{sec:append_setting}.

\begin{figure*}[!h]
\vskip -0.23in
\centering
\subfigure{
\begin{minipage}[t]{0.24\linewidth}
\centering
\includegraphics[width=\linewidth]{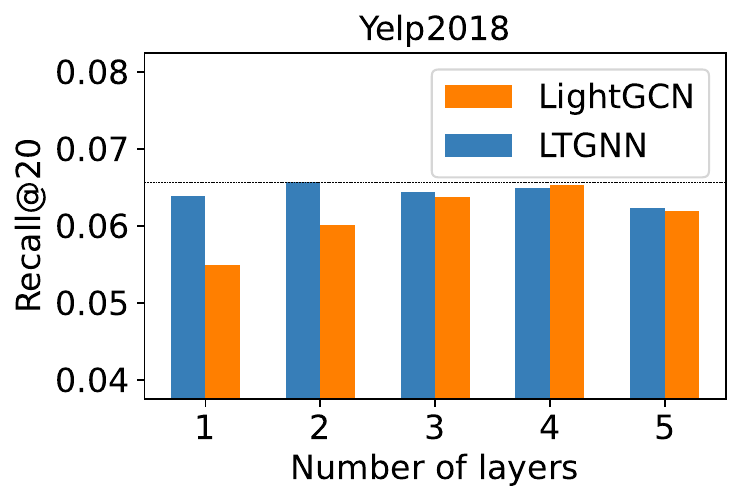}
\end{minipage}
}
\subfigure{
\begin{minipage}[t]{0.24\linewidth}
\centering
\includegraphics[width=\linewidth]{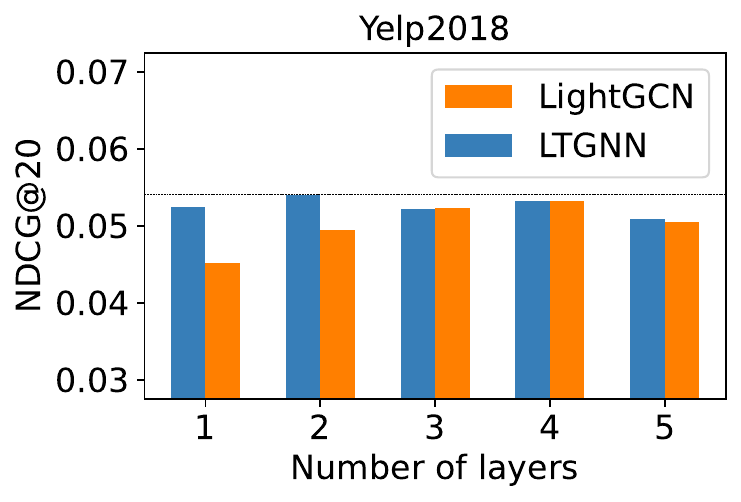}
\end{minipage}
\begin{minipage}[t]{0.24\linewidth}
\centering
\includegraphics[width=\linewidth]{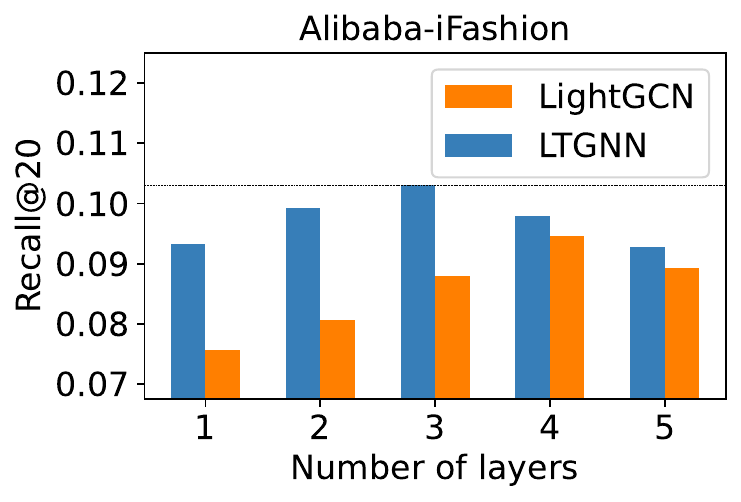}
\end{minipage}
}
\subfigure{
\begin{minipage}[t]{0.24\linewidth}
\centering
\includegraphics[width=\linewidth]{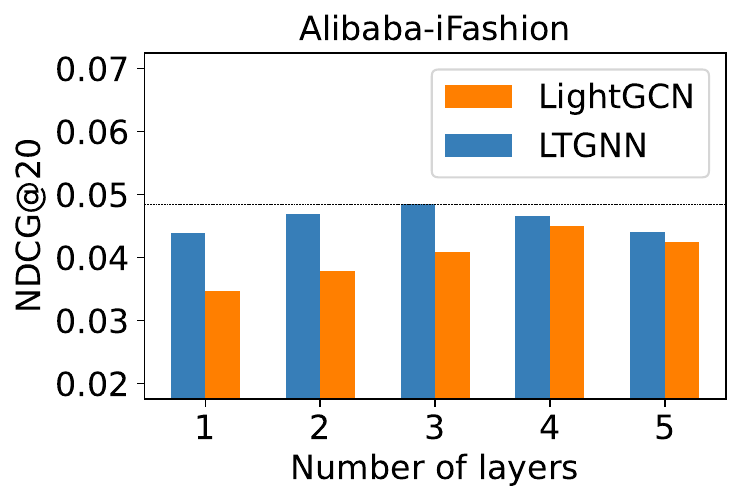}
\end{minipage}
}
\vskip -0.25in
\caption{Performance comparison between LTGNN and LightGCN using different layers on Yelp2018 and Alibaba-iFashion. }
\label{fig:perf_layer}
\vskip -0.25in
\end{figure*}

\begin{figure*}[!h]
\centering
\subfigure{
\begin{minipage}[t]{0.23\linewidth}
\centering
\includegraphics[width=\linewidth]{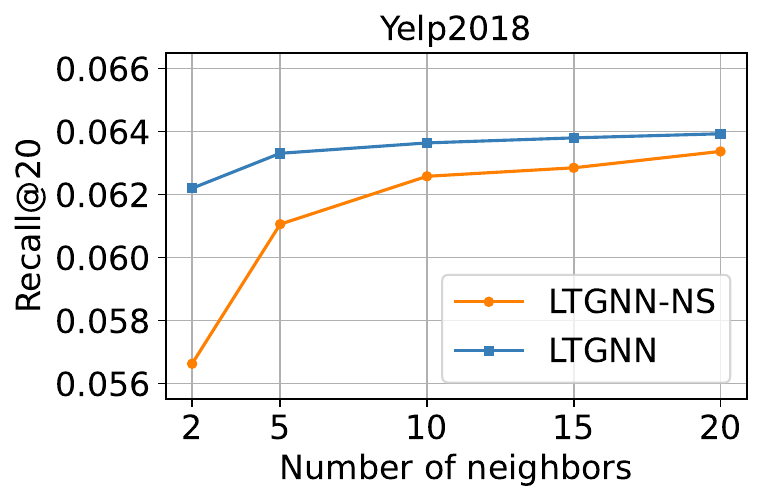}
\end{minipage}
}
\subfigure{
\begin{minipage}[t]{0.23\linewidth}
\centering
\includegraphics[width=\linewidth]{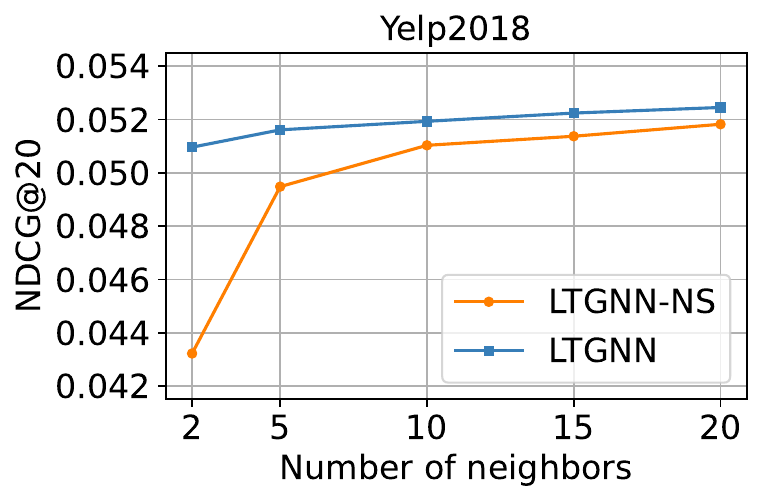}
\end{minipage}
}
\subfigure{
\begin{minipage}[t]{0.23\linewidth}
\centering
\includegraphics[width=\linewidth]{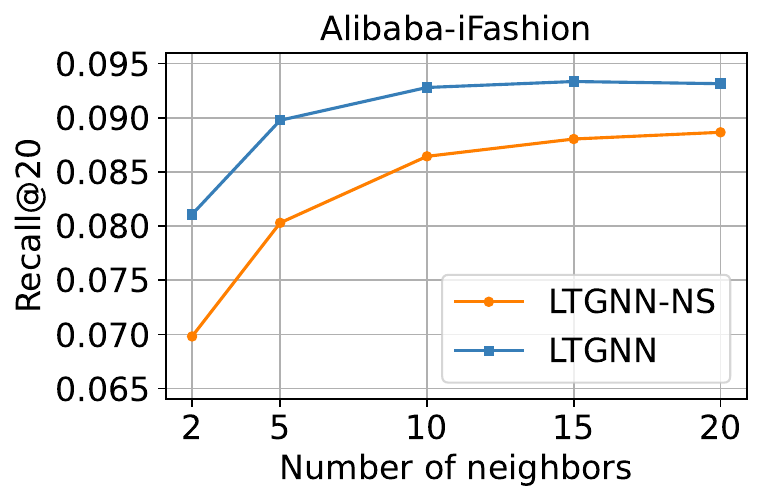}
\end{minipage}
}
\subfigure{
\begin{minipage}[t]{0.23\linewidth}
\centering
\includegraphics[width=\linewidth]{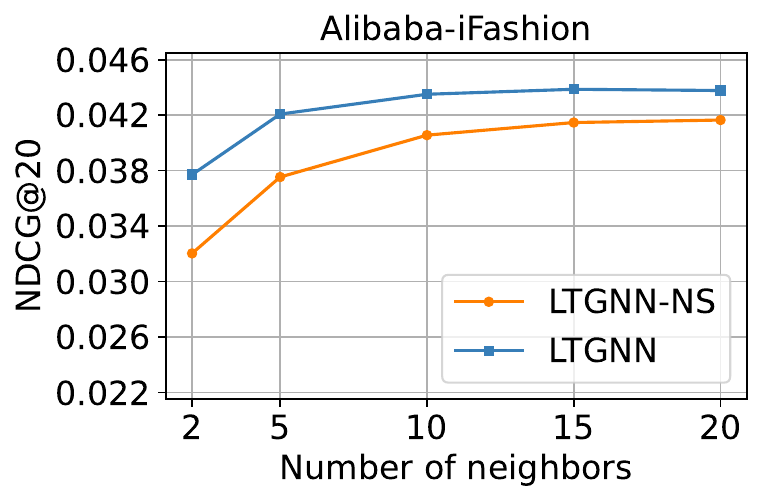}
\end{minipage}
}
\vskip -0.25in
\caption{Performance of a 1-layer LTGNN w.r.t different numbers of sampled neighbors on Yelp2018 and Alibaba-iFashion.}
\label{fig:neighbor_perf}
\vskip -0.15in
\end{figure*}

\subsection{Recommendation Performance}\label{sec:perf}

In this section, we mainly examine the recommendation performance of our proposed LTGNN, with a particular focus on comparing LTGNN with the most widely adopted GNN backbone LightGCN. 
We use out-of-memory (OOM) to indicate the methods that cannot run on the dataset due to memory limitations. The recommendation performance summarized in Table~\ref{tab:comparsion_perf} provides the following observations:
\vspace{-\topsep}
\begin{itemize}[leftmargin=0.2in]
    \item Our proposed LTGNN achieves comparable or better results on all three datasets compared to the strongest baselines. In particular, LTGNN outperforms all the baselines on Yelp and Alibaba-iFashion. The only exception is that the Recall@20 of LightGCN (L=3) outperforms LTGNN (L=1) on the Amazon-Large dataset. However, our NDCG@20 outperforms LightGCN (L=3), and LTGNN (L=1) is much more efficient compared with LightGCN (L=3), as LTGNN only uses one embedding propagation layer and very few randomly sampled neighbors. 
    
    \item The scalable variants of LightGCN improve the scalability of LightGCN by sacrificing its recommendation performance in most cases. For instance, the results for LightGCN-VR, LightGCN-NS, and LightGCN-GAS are much worse than LightGCN with full embedding propagation on Amazon-Large. In contrast, the proposed LTGNN has better efficiency than these variants and preserves the recommendation performance. 
    
    \item The performance of GNN-based methods like NGCF and LightGCN consistently outperforms earlier methods like MF and NCF. However, GNNs without scalability techniques can hardly be run large-scale datasets. 
    For instance, GC-MC, NGCF, and DGCF significantly outperform MF, but they are reported as OOM on the Amazon-Large dataset. 
    This suggests the necessity of pursuing scalable GNNs for improving the recommendation performance in large-scale industry scenarios. 
\end{itemize}
\vspace{-\topsep}

\begin{table}[t]
\vskip -0.05in
\caption{The comparison of running time on three datasets.}
\vskip -0.15in
\label{tab:comparsion_scale}
\begin{center}
\scalebox{0.72}
{
\begin{tabular}{l c|c|c|c}
\hline
\multicolumn{2}{c|}{\textbf{Dataset}} & \textbf{Yelp2018} & \textbf{Alibaba-iFashion} & \textbf{Amazon-Large} \\ \hline
\textbf{Method} & \textbf{\# Layer} & \textbf{Runnning Time} & \textbf{Runnning Time} & \textbf{Running Time}  \\ \hline\hline
LightGCN & \multirow{4}{*}{$L=3$} & 52.83s & 51.4s & 2999.35s \\ 
LightGCN-NS &                      & 46.09s &  51.70s & 4291.37s  \\ 
LightGCN-VR &                      &53.15s &  59.79s & 4849.59s \\ 
LightGCN-GAS &                     & 23.22s &  26.576s & 932.03s \\ \hline
LightGCN & \multirow{4}{*}{$L=2$} & 30.92s &  30.78s & 2061.75s \\ 
LightGCN-NS &                     & 37.17s &  26.89s & 1305.25s \\ 
LightGCN-VR &                     & 38.77s &  30.33s & 1545.34s  \\ 
LightGCN-GAS &                    & 22.92s &  25.04s & 903.78s \\ \hline
LightGCN & \multirow{4}{*}{$L=1$} & 16.95s &  18.02s & 1117.51s \\ 
LightGCN-NS &                     & 13.90s & 12.74s & 684.84s \\ 
LightGCN-VR &                     & 15.52s &  13.92s & 870.82s \\ 
LightGCN-GAS &                    & 14.53s &  13.35s & 729.22s \\ \hline
MF & - & 4.31s & 4.60s & 127.24s \\ \hline
{LTGNN} & \multirow{1}{*}{$L=1$}& {14.72s} & {13.68s} & {705.91s} \\ 
\hline
\end{tabular}
}
\end{center}
\vskip -0.2in
\end{table}

\subsection{Efficiency Analysis}\label{sec:scale}

To verify the scalability of LTGNN, we provide efficiency analysis in comparison with MF, LightGCN, and scalable variants of LightGCN with different layers on all three datasets: Yelp2018, Alibaba-iFasion, and Amazon-Large. From the running time shown in Table \ref{tab:comparsion_scale}, we draw the following conclusions:
\vspace{-\topsep}
\begin{itemize}[leftmargin=0.2in]
    \item Our proposed single-layer LTGNN achieves comparable running time with one-layer LightGCN with sampling, and outperforms the original LightGCN. This is consistent with our complexity analysis in Section \ref{sec:vr}. Moreover, LTGNN is faster than one-layer LightGCN with variance reduction, owing to our improved and efficient variance reduction (EVR) techniques. Although LTGNN is not substantially more efficient than some of the one-layer GNNs, it has much better recommendation performance, as shown in Figure~\ref{fig:perf_layer}. 
    
    \item LTGNN demonstrates significantly improved computational efficiency compared to baseline models with more than one layer. 

    When combined with the results from Table ~\ref{tab:comparsion_perf}, it becomes evident that LTGNN can maintain high performance while achieving a substantial enhancement in computational efficiency.

    \item While the running time of LTGNN is a few times longer than that of Matrix Factorization (MF) due to their constant factor difference in the complexity analysis (Table.~\ref{tab:complexity}), it's important to note that LTGNN already exhibits a nice and \textit{similar scaling behavior as MF}. This supports the better scalability of LTGNN in large-scale recommendations compared with other GNN-based methods. 

    \item An interesting observation is that on large-scale datasets, full-graph LightGCN surpasses LightGCN with neighbor sampling on efficiency. This is mainly because of the high CPU overhead of random sampling, which limits the utilization of GPUs. 
\end{itemize}
\vspace{-\topsep}
To further understand the efficiency of each component of LTGNN, a fine-grained efficiency analysis can be found in Appendix.~\ref{sec:append_exp}.

\subsection{Ablation Study}\label{sec:abl}

In this section, we provide extensive ablation studies to evaluate the effectiveness of different parts in our proposed framework. 
Extra experiments on the effect of hyperparameter $\alpha$ and different variance-reduction designs can be found in Appendix.~\ref{sec:append_exp}.

\noindent  \textbf{Effectivenss of implicit graph modeling.}
We conduct an ablation study to show the effect of embedding propagation layers and long-range collaborative signals. 
Particularly, we use the same setting for LightGCN and LTGNN and change the number of propagation layers $L$. 
As illustrated in Figure~\ref{fig:perf_layer}, we have two key observations: 1) LTGNN with only 1 or 2 propagation layers can reach better performance in comparison with LightGCN with more than 3 layers, which demonstrates the strong long-range modeling capability of our proposed model; 2) Adding more propagation layers into LTGNN will not significantly improve its performance, which means $L=1$ or $L=2$ are the best choices for LTGNN to balance its performance and scalability.  

\noindent  \textbf{Effectiveness of efficient variance reduction.} 
In this study, we aim to demonstrate the effectiveness of our proposed EVR algorithm by showing the impact of the number of neighbors on recommendation performance. 
As shown in Figure~\ref{fig:neighbor_perf}, LTGNN with efficient variance reduction consistently outperforms its vanilla neighbor sampling variant (i.e., LTGNN-NS) regardless of the number of neighbors, illustrating its effect in reducing the large approximation error caused by neighbor sampling. 
The recommendation performance of LTGNN with efficient variance reduction is remarkably stable, even under extreme conditions like sampling only 2 neighbors for each target node. 
This indicates the great potential of our proposed LTGNN in large-scale recommendations, as a GNN with only one propagation layer and two random neighbors will be ultra-efficient compared with previous designs that incur a large number of layers and neighbors. 

\begin{figure}[t]
\centering
\vskip -0.16in
\subfigure{
\begin{minipage}[t]{0.47\linewidth}
\centering
\includegraphics[width=\linewidth]{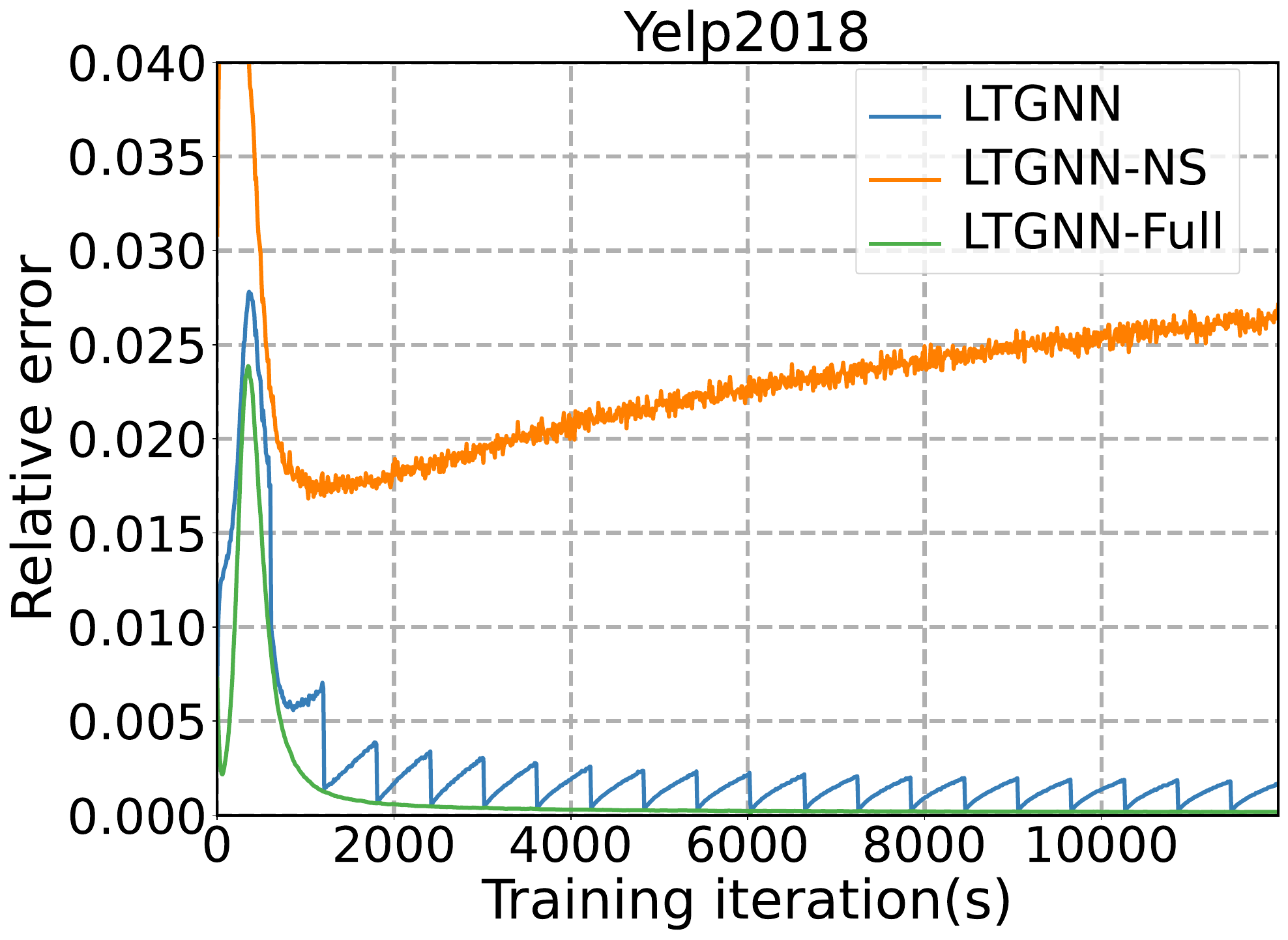}
\end{minipage}
}
\subfigure{
\begin{minipage}[t]{0.45\linewidth}
\centering
\includegraphics[width=\linewidth]{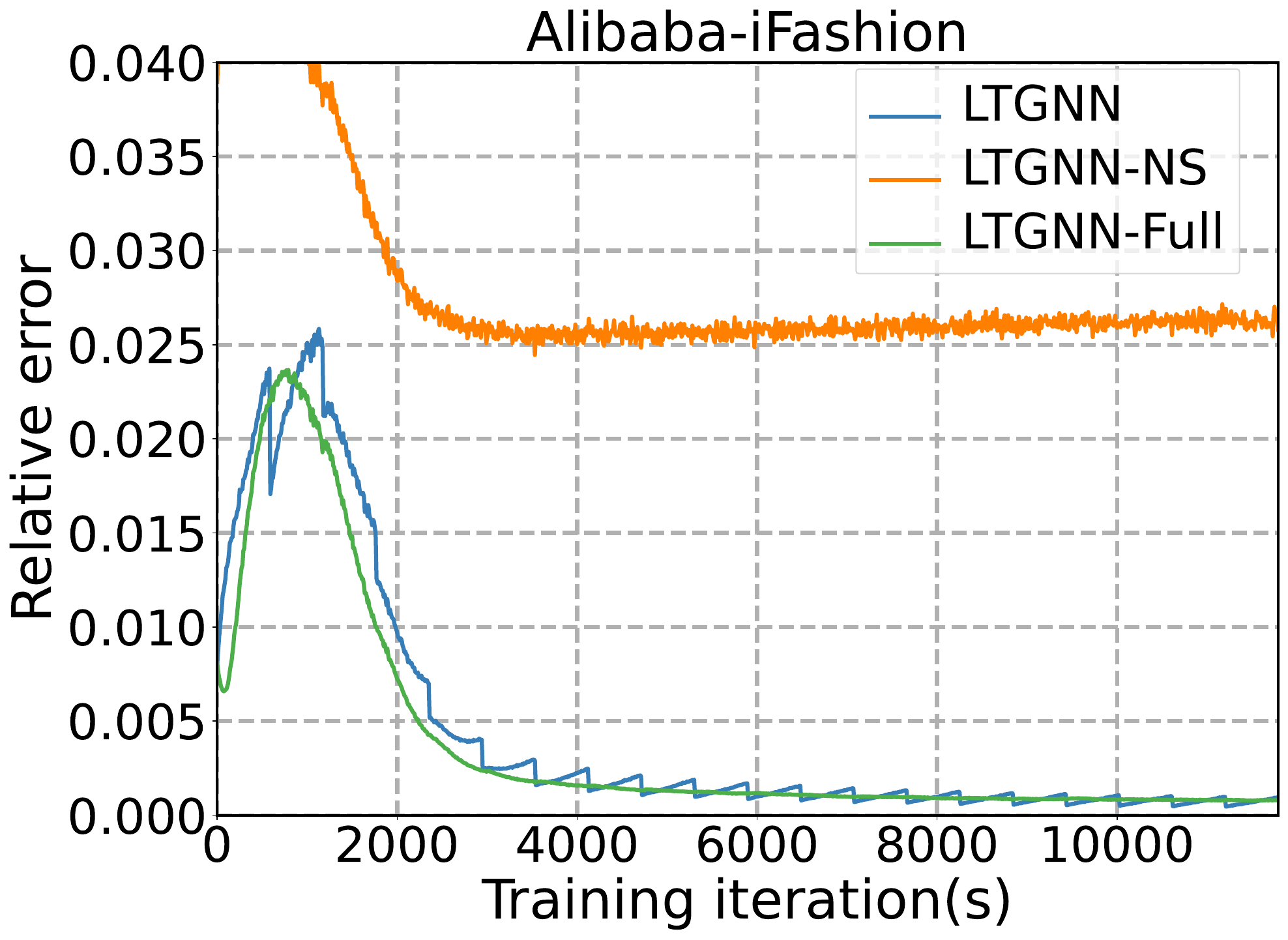}
\end{minipage}
}
\vskip -0.22in
\caption{The relative error between the model output $\boldsymbol{E}^k_{out}$ and the exact PPNP propagation result $\boldsymbol{E}^k_{PPNP}$ of the embeddings (i.e., $||\boldsymbol{E}^k_{out} - \boldsymbol{E}^k_{PPNP}||_F / ||\boldsymbol{E}^k_{PPNP}||_F$).}
\vspace{-0.2in}
\label{fig:num_anal}
\end{figure}

\noindent  \textbf{Numerical Analysis.}
In this experiment, we compute the PPNP embedding propagation result $\boldsymbol{E}^k_{PPNP}$ for the target nodes as an indicator of long-range modeling ability , and we compute the relative error between the model output $\boldsymbol{E}^k_L$ and the PPNP computation result. 
We use $L=1$ for LTGNN and its two inferior variants without efficient variance reduction - LTGNN-NS and LTGNN-Full, which denotes LTGNN with random neighbor sampling and LTGNN with exact full neighbor aggregation. 
From Figure~\ref{fig:num_anal}, we have two observations as follows: 1) On both datasets, the output of LTGNN converges to PPNP after around 4000 training iterations (i.e., less than 10 training epochs), which means our proposed LTGNN can capture the long-range dependencies in user-item graphs by using only one propagation layer; 2) By comparing LTGNN with its variants, it is obvious that neighbor sampling without variance reduction seriously hurts the modeling ability of LTGNN, and our proposed LTGNN has similar convergence curves in comparison to LTGNN with full aggregation, showing the effectiveness of our proposed efficient variance reduction method.

\section{Related Work}
In this section, we summarize the related works on graph-based collaborative filtering and scalable GNNs.

\subsection{Graph Collaborative Filtering Models for Recommendations}
In modern recommender systems, collaborative filtering (CF) is one of the most representative paradigm~\citep{fan2023untargeted,fan2020graph,fan2022graph} to understand users' preferences. 
The basic idea of CF is to decompose user-item interaction data into trainable user and item embeddings, and then reconstruct the missing interactions~\citep{koren2008factorization, he2017neural, fan2019deep, fan2023adversarial}.
Early works in CF mainly model the user-item interactions with scaled dot-product~\cite{koren2008factorization,fan2022graph,chen2023fairly}, MLPs~\cite{he2017neural, fan2018deep}, and LSTMs~\cite{fan2019deep, guo2020attentional}. 
However, these models fail to model the high-order collaborative signals between users and items, leading to sub-optimal representations of users and items.

In recent years, a promising line of studies has incorporated GNNs into CF-based recommender systems. 
The key advantage of utilizing GNN-based recommendation models is that GNNs can easily capture long-range dependencies via the information propagation mechanism on the user-item graph. 
For example, an early exploration, GC-MC, completes the missing user-item interactions with graph convolutional autoencoders~\citep{berg2017graph}. 
For large-scale recommendation scenarios, PinSAGE~\citep{ying2018graph} adapts the key idea of GraphSAGE~\citep{hamilton2017inductive} to recommendations and achieves promising results. 
Another significant milestone in GNN-based recommendations is the NGCF~\citep{wang2019neural}, which explicitly formulates the concept of collaborative signals and models them as high-order connectivities by message-passing propagations. 
Afterward, LightGCN~\citep{he2020lightgcn} indicates the non-linear activation functions and feature transformations in GNN-based recommender systems are redundant, and proposes to simplify existing GNNs while achieving promising performance. 
However, despite the success of previous GNN-based models in capturing user preferences, existing works fail to address
the neighborhood explosion problem on large-scale recommendation scenarios, which indicates that the scalability of GNN-based recommender systems remains an open question. 

\subsection{Scalability of Graph Neural Networks}

Recently, extensive literature has explored the scalability of GNNs on large-scale graphs~\cite{chen2024macro,wang2023fast}, with a wide range of research focusing on sampling methods, pre-computing methods, post-computing methods, and memory-based methods. 
Sampling-based methods lower the computation and memory requirements of GNNs by using a mini-batch training strategy on GNNs, which samples a limited number of neighbors for target nodes in a node-wise~\cite{hamilton2017inductive, chen2018stochastic, cong2020minimal}, layer-wise~\cite{chen2018fastgcn, zou2019layer}, or subgraph-wise~\cite{zeng2019graphsaint} manner, mitigating the well-known neighborhood explosion problem. 
However, sampling-based methods inevitably omit a large number of neighbors for aggregation, resulting in large random errors. 
As a remedy for this limitation, memory-based methods~\cite{fey2021gnnautoscale, yu2022graphfm, xue2023lazygnn} leverage the historical feature memories to complement the missing information of out-of-batch nodes, approximating the full aggregation outputs while enjoying the efficiency of only updating the in-batch nodes. 
Unfortunately, these methods may have complexities similar to full neighbor aggregations, which block the way to linear complexity. 
Besides, pre-computing or post-computing methods decouple feature transformation and feature aggregations, enabling capturing long-range dependencies in graphs while only training a feature transformation model. 
In particular, pre-computing methods pre-calculate the feature aggregation result before training~\citep{wu2019simplifying, zhang2022graph}, while post-computing methods firstly train a feature transformation model and leverage unsupervised algorithms, such as label propagation, to predict the labels~\citep{huang2020combining, zhu2005semi}.
In retrospect, pre-computing/post-computing methods may sacrifice the advantage of end-to-end training, and the approximation error of these methods is still unsatisfactory. 
These limitations in existing scalable GNNs strongly necessitate our pursuit of scalable GNNs for large-scale recommender systems in this paper.

\section{Conclusion} 
Scalability is a major challenge for GNN-based recommender systems, as they often require many computational resources to handle large-scale user-item interaction data. 
To address this challenge, we propose a novel scalable GNN model for recommendation, which leverages implicit graph modeling and variance-reduced neighbor sampling to capture long-range collaborative signals while achieving a desirable linear complexity. 
The proposed LTGNN only needs one propagation layer and a small number of one-hop neighbors, which reduces the computation complexity to be linear to the number of edges, showing great potential in industrial recommendation applications. 
Extensive experiments on three real-world datasets are conducted to demonstrate the effectiveness and efficiency of our proposed model. 
For future work, we plan to extend our design to more recommendation tasks, such as CTR prediction, and explore the deployment of LTGNN in industrial recommender systems.

\begin{acks}
The research described in this paper has been partly supported by NSFC (project no. 62102335), General Research Funds from the Hong Kong Research Grants Council (project no. PolyU 15200021, 15207322, and 15200023), internal research funds from The Hong Kong Polytechnic University (project no. P0036200, P0042693, P0048\\625, P0048752), Research Collaborative Project no. P0041282, and SHTM Interdisciplinary Large Grant (project no. P0043302). 
Xiaorui Liu is partially supported by the Amazon Research Award. 
\end{acks}

\balance
\bibliographystyle{ACM-Reference-Format}
\bibliography{ref_short.bib}


\begin{thebibliography}{55}


\ifx \showCODEN    \undefined \def \showCODEN     #1{\unskip}     \fi
\ifx \showDOI      \undefined \def \showDOI       #1{#1}\fi
\ifx \showISBNx    \undefined \def \showISBNx     #1{\unskip}     \fi
\ifx \showISBNxiii \undefined \def \showISBNxiii  #1{\unskip}     \fi
\ifx \showISSN     \undefined \def \showISSN      #1{\unskip}     \fi
\ifx \showLCCN     \undefined \def \showLCCN      #1{\unskip}     \fi
\ifx \shownote     \undefined \def \shownote      #1{#1}          \fi
\ifx \showarticletitle \undefined \def \showarticletitle #1{#1}   \fi
\ifx \showURL      \undefined \def \showURL       {\relax}        \fi
\providecommand\bibfield[2]{#2}
\providecommand\bibinfo[2]{#2}
\providecommand\natexlab[1]{#1}
\providecommand\showeprint[2][]{arXiv:#2}

\bibitem[Berg et~al\mbox{.}(2017)]%
        {berg2017graph}
\bibfield{author}{\bibinfo{person}{Rianne van~den Berg}, \bibinfo{person}{Thomas~N Kipf}, {and} \bibinfo{person}{Max Welling}.} \bibinfo{year}{2017}\natexlab{}.
\newblock \showarticletitle{Graph convolutional matrix completion}.
\newblock \bibinfo{journal}{\emph{arXiv preprint arXiv:1706.02263}} (\bibinfo{year}{2017}).
\newblock


\bibitem[Chen et~al\mbox{.}(2024)]%
        {chen2024macro}
\bibfield{author}{\bibinfo{person}{Hao Chen}, \bibinfo{person}{Yuanchen Bei}, \bibinfo{person}{Qijie Shen}, \bibinfo{person}{Yue Xu}, \bibinfo{person}{Sheng Zhou}, \bibinfo{person}{Wenbing Huang}, \bibinfo{person}{Feiran Huang}, \bibinfo{person}{Senzhang Wang}, {and} \bibinfo{person}{Xiao Huang}.} \bibinfo{year}{2024}\natexlab{}.
\newblock \showarticletitle{Macro Graph Neural Networks for Online Billion-Scale Recommender Systems}. In \bibinfo{booktitle}{\emph{WWW}}.
\newblock


\bibitem[Chen et~al\mbox{.}(2018a)]%
        {chen2018fastgcn}
\bibfield{author}{\bibinfo{person}{Jie Chen}, \bibinfo{person}{Tengfei Ma}, {and} \bibinfo{person}{Cao Xiao}.} \bibinfo{year}{2018}\natexlab{a}.
\newblock \showarticletitle{Fastgcn: fast learning with graph convolutional networks via importance sampling}. In \bibinfo{booktitle}{\emph{ICLR}}.
\newblock


\bibitem[Chen et~al\mbox{.}(2018b)]%
        {chen2018stochastic}
\bibfield{author}{\bibinfo{person}{Jianfei Chen}, \bibinfo{person}{Jun Zhu}, {and} \bibinfo{person}{Le Song}.} \bibinfo{year}{2018}\natexlab{b}.
\newblock \showarticletitle{Stochastic Training of Graph Convolutional Networks with Variance Reduction}. In \bibinfo{booktitle}{\emph{ICML}}.
\newblock


\bibitem[Chen et~al\mbox{.}(2023)]%
        {chen2023fairly}
\bibfield{author}{\bibinfo{person}{Xiao Chen}, \bibinfo{person}{Wenqi Fan}, \bibinfo{person}{Jingfan Chen}, \bibinfo{person}{Haochen Liu}, \bibinfo{person}{Zitao Liu}, \bibinfo{person}{Zhaoxiang Zhang}, {and} \bibinfo{person}{Qing Li}.} \bibinfo{year}{2023}\natexlab{}.
\newblock \showarticletitle{Fairly adaptive negative sampling for recommendations}. In \bibinfo{booktitle}{\emph{WWW}}.
\newblock


\bibitem[Cong et~al\mbox{.}(2020)]%
        {cong2020minimal}
\bibfield{author}{\bibinfo{person}{Weilin Cong}, \bibinfo{person}{Rana Forsati}, \bibinfo{person}{Mahmut Kandemir}, {and} \bibinfo{person}{Mehrdad Mahdavi}.} \bibinfo{year}{2020}\natexlab{}.
\newblock \showarticletitle{Minimal variance sampling with provable guarantees for fast training of graph neural networks}. In \bibinfo{booktitle}{\emph{KDD}}.
\newblock


\bibitem[Covington et~al\mbox{.}(2016)]%
        {covington2016deep}
\bibfield{author}{\bibinfo{person}{Paul Covington}, \bibinfo{person}{Jay Adams}, {and} \bibinfo{person}{Emre Sargin}.} \bibinfo{year}{2016}\natexlab{}.
\newblock \showarticletitle{Deep neural networks for youtube recommendations}. In \bibinfo{booktitle}{\emph{RecSys}}.
\newblock


\bibitem[Duan et~al\mbox{.}(2022)]%
        {duan2022comprehensive}
\bibfield{author}{\bibinfo{person}{Keyu Duan}, \bibinfo{person}{Zirui Liu}, \bibinfo{person}{Peihao Wang}, \bibinfo{person}{Wenqing Zheng}, \bibinfo{person}{Kaixiong Zhou}, \bibinfo{person}{Tianlong Chen}, \bibinfo{person}{Xia Hu}, {and} \bibinfo{person}{Zhangyang Wang}.} \bibinfo{year}{2022}\natexlab{}.
\newblock \showarticletitle{A comprehensive study on large-scale graph training: Benchmarking and rethinking}.
\newblock \bibinfo{journal}{\emph{NeurIPS}} (\bibinfo{year}{2022}).
\newblock


\bibitem[El~Ghaoui et~al\mbox{.}(2021)]%
        {el2021implicit}
\bibfield{author}{\bibinfo{person}{Laurent El~Ghaoui}, \bibinfo{person}{Fangda Gu}, \bibinfo{person}{Bertrand Travacca}, \bibinfo{person}{Armin Askari}, {and} \bibinfo{person}{Alicia Tsai}.} \bibinfo{year}{2021}\natexlab{}.
\newblock \showarticletitle{Implicit deep learning}.
\newblock \bibinfo{journal}{\emph{SIAM Journal on Mathematics of Data Science}} \bibinfo{volume}{3}, \bibinfo{number}{3} (\bibinfo{year}{2021}), \bibinfo{pages}{930--958}.
\newblock


\bibitem[Elkahky et~al\mbox{.}(2015)]%
        {elkahky2015multi}
\bibfield{author}{\bibinfo{person}{Ali~Mamdouh Elkahky}, \bibinfo{person}{Yang Song}, {and} \bibinfo{person}{Xiaodong He}.} \bibinfo{year}{2015}\natexlab{}.
\newblock \showarticletitle{A multi-view deep learning approach for cross domain user modeling in recommendation systems}. In \bibinfo{booktitle}{\emph{WWW}}.
\newblock


\bibitem[Fan et~al\mbox{.}(2018)]%
        {fan2018deep}
\bibfield{author}{\bibinfo{person}{Wenqi Fan}, \bibinfo{person}{Qing Li}, {and} \bibinfo{person}{Min Cheng}.} \bibinfo{year}{2018}\natexlab{}.
\newblock \showarticletitle{Deep modeling of social relations for recommendation}. In \bibinfo{booktitle}{\emph{AAAI}}.
\newblock


\bibitem[Fan et~al\mbox{.}(2022)]%
        {fan2022graph}
\bibfield{author}{\bibinfo{person}{Wenqi Fan}, \bibinfo{person}{Xiaorui Liu}, \bibinfo{person}{Wei Jin}, \bibinfo{person}{Xiangyu Zhao}, \bibinfo{person}{Jiliang Tang}, {and} \bibinfo{person}{Qing Li}.} \bibinfo{year}{2022}\natexlab{}.
\newblock \showarticletitle{Graph trend filtering networks for recommendation}. In \bibinfo{booktitle}{\emph{SIGIR}}.
\newblock


\bibitem[Fan et~al\mbox{.}(2019a)]%
        {fan2019graph}
\bibfield{author}{\bibinfo{person}{Wenqi Fan}, \bibinfo{person}{Yao Ma}, \bibinfo{person}{Qing Li}, \bibinfo{person}{Yuan He}, \bibinfo{person}{Eric Zhao}, \bibinfo{person}{Jiliang Tang}, {and} \bibinfo{person}{Dawei Yin}.} \bibinfo{year}{2019}\natexlab{a}.
\newblock \showarticletitle{Graph neural networks for social recommendation}. In \bibinfo{booktitle}{\emph{WWW}}.
\newblock


\bibitem[Fan et~al\mbox{.}(2020)]%
        {fan2020graph}
\bibfield{author}{\bibinfo{person}{Wenqi Fan}, \bibinfo{person}{Yao Ma}, \bibinfo{person}{Qing Li}, \bibinfo{person}{Jianping Wang}, \bibinfo{person}{Guoyong Cai}, \bibinfo{person}{Jiliang Tang}, {and} \bibinfo{person}{Dawei Yin}.} \bibinfo{year}{2020}\natexlab{}.
\newblock \showarticletitle{A graph neural network framework for social recommendations}.
\newblock \bibinfo{journal}{\emph{TKDE}} \bibinfo{volume}{34}, \bibinfo{number}{5} (\bibinfo{year}{2020}), \bibinfo{pages}{2033--2047}.
\newblock


\bibitem[Fan et~al\mbox{.}(2019b)]%
        {fan2019deep}
\bibfield{author}{\bibinfo{person}{Wenqi Fan}, \bibinfo{person}{Yao Ma}, \bibinfo{person}{Dawei Yin}, \bibinfo{person}{Jianping Wang}, \bibinfo{person}{Jiliang Tang}, {and} \bibinfo{person}{Qing Li}.} \bibinfo{year}{2019}\natexlab{b}.
\newblock \showarticletitle{Deep social collaborative filtering}. In \bibinfo{booktitle}{\emph{RecSys}}.
\newblock


\bibitem[Fan et~al\mbox{.}(2023a)]%
        {fan2023untargeted}
\bibfield{author}{\bibinfo{person}{Wenqi Fan}, \bibinfo{person}{Shijie Wang}, \bibinfo{person}{Xiao-yong Wei}, \bibinfo{person}{Xiaowei Mei}, {and} \bibinfo{person}{Qing Li}.} \bibinfo{year}{2023}\natexlab{a}.
\newblock \showarticletitle{Untargeted Black-box Attacks for Social Recommendations}.
\newblock \bibinfo{journal}{\emph{arXiv preprint arXiv:2311.07127}} (\bibinfo{year}{2023}).
\newblock


\bibitem[Fan et~al\mbox{.}(2023b)]%
        {fan2023adversarial}
\bibfield{author}{\bibinfo{person}{Wenqi Fan}, \bibinfo{person}{Xiangyu Zhao}, \bibinfo{person}{Qing Li}, \bibinfo{person}{Tyler Derr}, \bibinfo{person}{Yao Ma}, \bibinfo{person}{Hui Liu}, \bibinfo{person}{Jianping Wang}, {and} \bibinfo{person}{Jiliang Tang}.} \bibinfo{year}{2023}\natexlab{b}.
\newblock \showarticletitle{Adversarial Attacks for Black-Box Recommender Systems Via Copying Transferable Cross-Domain User Profiles}.
\newblock \bibinfo{journal}{\emph{TKDE}} (\bibinfo{year}{2023}).
\newblock


\bibitem[Fey et~al\mbox{.}(2021)]%
        {fey2021gnnautoscale}
\bibfield{author}{\bibinfo{person}{Matthias Fey}, \bibinfo{person}{Jan~E Lenssen}, \bibinfo{person}{Frank Weichert}, {and} \bibinfo{person}{Jure Leskovec}.} \bibinfo{year}{2021}\natexlab{}.
\newblock \showarticletitle{Gnnautoscale: Scalable and expressive graph neural networks via historical embeddings}. In \bibinfo{booktitle}{\emph{ICML}}.
\newblock


\bibitem[Gasteiger et~al\mbox{.}(2018)]%
        {gasteiger2018predict}
\bibfield{author}{\bibinfo{person}{Johannes Gasteiger}, \bibinfo{person}{Aleksandar Bojchevski}, {and} \bibinfo{person}{Stephan G{\"u}nnemann}.} \bibinfo{year}{2018}\natexlab{}.
\newblock \showarticletitle{Predict then Propagate: Graph Neural Networks meet Personalized PageRank}. In \bibinfo{booktitle}{\emph{ICLR}}.
\newblock


\bibitem[Gu et~al\mbox{.}(2020)]%
        {gu2020implicit}
\bibfield{author}{\bibinfo{person}{Fangda Gu}, \bibinfo{person}{Heng Chang}, \bibinfo{person}{Wenwu Zhu}, \bibinfo{person}{Somayeh Sojoudi}, {and} \bibinfo{person}{Laurent El~Ghaoui}.} \bibinfo{year}{2020}\natexlab{}.
\newblock \showarticletitle{Implicit graph neural networks}. In \bibinfo{booktitle}{\emph{NeurIPS}}.
\newblock


\bibitem[Guo et~al\mbox{.}(2020)]%
        {guo2020attentional}
\bibfield{author}{\bibinfo{person}{Qing Guo}, \bibinfo{person}{Zhu Sun}, \bibinfo{person}{Jie Zhang}, {and} \bibinfo{person}{Yin-Leng Theng}.} \bibinfo{year}{2020}\natexlab{}.
\newblock \showarticletitle{An attentional recurrent neural network for personalized next location recommendation}. In \bibinfo{booktitle}{\emph{AAAI}}.
\newblock


\bibitem[Hamilton et~al\mbox{.}(2017)]%
        {hamilton2017inductive}
\bibfield{author}{\bibinfo{person}{Will Hamilton}, \bibinfo{person}{Zhitao Ying}, {and} \bibinfo{person}{Jure Leskovec}.} \bibinfo{year}{2017}\natexlab{}.
\newblock \showarticletitle{Inductive representation learning on large graphs}. In \bibinfo{booktitle}{\emph{NeurIPS}}.
\newblock


\bibitem[He et~al\mbox{.}(2020)]%
        {he2020lightgcn}
\bibfield{author}{\bibinfo{person}{Xiangnan He}, \bibinfo{person}{Kuan Deng}, \bibinfo{person}{Xiang Wang}, \bibinfo{person}{Yan Li}, \bibinfo{person}{Yongdong Zhang}, {and} \bibinfo{person}{Meng Wang}.} \bibinfo{year}{2020}\natexlab{}.
\newblock \showarticletitle{Lightgcn: Simplifying and powering graph convolution network for recommendation}. In \bibinfo{booktitle}{\emph{SIGIR}}.
\newblock


\bibitem[He et~al\mbox{.}(2018)]%
        {he2018nais}
\bibfield{author}{\bibinfo{person}{Xiangnan He}, \bibinfo{person}{Zhankui He}, \bibinfo{person}{Jingkuan Song}, \bibinfo{person}{Zhenguang Liu}, \bibinfo{person}{Yu-Gang Jiang}, {and} \bibinfo{person}{Tat-Seng Chua}.} \bibinfo{year}{2018}\natexlab{}.
\newblock \showarticletitle{NAIS: Neural attentive item similarity model for recommendation}.
\newblock \bibinfo{journal}{\emph{TKDE}} \bibinfo{volume}{30}, \bibinfo{number}{12} (\bibinfo{year}{2018}), \bibinfo{pages}{2354--2366}.
\newblock


\bibitem[He et~al\mbox{.}(2017)]%
        {he2017neural}
\bibfield{author}{\bibinfo{person}{Xiangnan He}, \bibinfo{person}{Lizi Liao}, \bibinfo{person}{Hanwang Zhang}, \bibinfo{person}{Liqiang Nie}, \bibinfo{person}{Xia Hu}, {and} \bibinfo{person}{Tat-Seng Chua}.} \bibinfo{year}{2017}\natexlab{}.
\newblock \showarticletitle{Neural collaborative filtering}. In \bibinfo{booktitle}{\emph{WWW}}.
\newblock


\bibitem[Huang et~al\mbox{.}(2020)]%
        {huang2020combining}
\bibfield{author}{\bibinfo{person}{Qian Huang}, \bibinfo{person}{Horace He}, \bibinfo{person}{Abhay Singh}, \bibinfo{person}{Ser-Nam Lim}, {and} \bibinfo{person}{Austin Benson}.} \bibinfo{year}{2020}\natexlab{}.
\newblock \showarticletitle{Combining Label Propagation and Simple Models out-performs Graph Neural Networks}. In \bibinfo{booktitle}{\emph{ICLR}}.
\newblock


\bibitem[Huang et~al\mbox{.}(2021)]%
        {huang2021mixgcf}
\bibfield{author}{\bibinfo{person}{Tinglin Huang}, \bibinfo{person}{Yuxiao Dong}, \bibinfo{person}{Ming Ding}, \bibinfo{person}{Zhen Yang}, \bibinfo{person}{Wenzheng Feng}, \bibinfo{person}{Xinyu Wang}, {and} \bibinfo{person}{Jie Tang}.} \bibinfo{year}{2021}\natexlab{}.
\newblock \showarticletitle{Mixgcf: An improved training method for graph neural network-based recommender systems}. In \bibinfo{booktitle}{\emph{KDD}}.
\newblock


\bibitem[Jin et~al\mbox{.}(2023)]%
        {jin2023amazon}
\bibfield{author}{\bibinfo{person}{Wei Jin}, \bibinfo{person}{Haitao Mao}, \bibinfo{person}{Zheng Li}, \bibinfo{person}{Haoming Jiang}, \bibinfo{person}{Chen Luo}, \bibinfo{person}{Hongzhi Wen}, \bibinfo{person}{Haoyu Han}, \bibinfo{person}{Hanqing Lu}, \bibinfo{person}{Zhengyang Wang}, \bibinfo{person}{Ruirui Li}, {et~al\mbox{.}}} \bibinfo{year}{2023}\natexlab{}.
\newblock \showarticletitle{Amazon-M2: A Multilingual Multi-locale Shopping Session Dataset for Recommendation and Text Generation}. In \bibinfo{booktitle}{\emph{NeurIPS}}.
\newblock


\bibitem[Kingma and Ba(2015)]%
        {kingma2014adam}
\bibfield{author}{\bibinfo{person}{Diederik~P Kingma} {and} \bibinfo{person}{Jimmy Ba}.} \bibinfo{year}{2015}\natexlab{}.
\newblock \showarticletitle{Adam: A method for stochastic optimization}. In \bibinfo{booktitle}{\emph{ICLR}}.
\newblock


\bibitem[Kipf and Welling(2017)]%
        {kipf2016semi}
\bibfield{author}{\bibinfo{person}{Thomas~N Kipf} {and} \bibinfo{person}{Max Welling}.} \bibinfo{year}{2017}\natexlab{}.
\newblock \showarticletitle{Semi-supervised classification with graph convolutional networks}. In \bibinfo{booktitle}{\emph{ICLR}}.
\newblock


\bibitem[Koren(2008)]%
        {koren2008factorization}
\bibfield{author}{\bibinfo{person}{Yehuda Koren}.} \bibinfo{year}{2008}\natexlab{}.
\newblock \showarticletitle{Factorization meets the neighborhood: a multifaceted collaborative filtering model}. In \bibinfo{booktitle}{\emph{KDD}}.
\newblock


\bibitem[Lawrence(1998)]%
        {lawrence1998pagerank}
\bibfield{author}{\bibinfo{person}{Page Lawrence}.} \bibinfo{year}{1998}\natexlab{}.
\newblock \showarticletitle{The pagerank citation ranking: Bringing order to the web}.
\newblock \bibinfo{journal}{\emph{Technical report}} (\bibinfo{year}{1998}).
\newblock


\bibitem[Li et~al\mbox{.}(2022)]%
        {li2022unbiased}
\bibfield{author}{\bibinfo{person}{Mingjie Li}, \bibinfo{person}{Yifei Wang}, \bibinfo{person}{Yisen Wang}, {and} \bibinfo{person}{Zhouchen Lin}.} \bibinfo{year}{2022}\natexlab{}.
\newblock \showarticletitle{Unbiased Stochastic Proximal Solver for Graph Neural Networks with Equilibrium States}. In \bibinfo{booktitle}{\emph{ICLR}}.
\newblock


\bibitem[Li et~al\mbox{.}(2020)]%
        {li2020hierarchical}
\bibfield{author}{\bibinfo{person}{Zhao Li}, \bibinfo{person}{Xin Shen}, \bibinfo{person}{Yuhang Jiao}, \bibinfo{person}{Xuming Pan}, \bibinfo{person}{Pengcheng Zou}, \bibinfo{person}{Xianling Meng}, \bibinfo{person}{Chengwei Yao}, {and} \bibinfo{person}{Jiajun Bu}.} \bibinfo{year}{2020}\natexlab{}.
\newblock \showarticletitle{Hierarchical bipartite graph neural networks: Towards large-scale e-commerce applications}. In \bibinfo{booktitle}{\emph{ICDE}}.
\newblock


\bibitem[Liu et~al\mbox{.}(2023)]%
        {liu2023generative}
\bibfield{author}{\bibinfo{person}{Chengyi Liu}, \bibinfo{person}{Wenqi Fan}, \bibinfo{person}{Yunqing Liu}, \bibinfo{person}{Jiatong Li}, \bibinfo{person}{Hang Li}, \bibinfo{person}{Hui Liu}, \bibinfo{person}{Jiliang Tang}, {and} \bibinfo{person}{Qing Li}.} \bibinfo{year}{2023}\natexlab{}.
\newblock \showarticletitle{Generative diffusion models on graphs: Methods and applications}. In \bibinfo{booktitle}{\emph{IJCAI}}.
\newblock


\bibitem[Mao et~al\mbox{.}(2021)]%
        {mao2021simplex}
\bibfield{author}{\bibinfo{person}{Kelong Mao}, \bibinfo{person}{Jieming Zhu}, \bibinfo{person}{Jinpeng Wang}, \bibinfo{person}{Quanyu Dai}, \bibinfo{person}{Zhenhua Dong}, \bibinfo{person}{Xi Xiao}, {and} \bibinfo{person}{Xiuqiang He}.} \bibinfo{year}{2021}\natexlab{}.
\newblock \showarticletitle{SimpleX: A simple and strong baseline for collaborative filtering}. In \bibinfo{booktitle}{\emph{CIKM}}.
\newblock


\bibitem[Rendle et~al\mbox{.}(2009)]%
        {rendle2009bpr}
\bibfield{author}{\bibinfo{person}{Steffen Rendle}, \bibinfo{person}{Christoph Freudenthaler}, \bibinfo{person}{Zeno Gantner}, {and} \bibinfo{person}{Lars Schmidt-Thieme}.} \bibinfo{year}{2009}\natexlab{}.
\newblock \showarticletitle{BPR: Bayesian personalized ranking from implicit feedback}. In \bibinfo{booktitle}{\emph{UAI}}.
\newblock


\bibitem[Teng et~al\mbox{.}(2016)]%
        {teng2016scalable}
\bibfield{author}{\bibinfo{person}{Shang-Hua Teng} {et~al\mbox{.}}} \bibinfo{year}{2016}\natexlab{}.
\newblock \showarticletitle{Scalable algorithms for data and network analysis}.
\newblock \bibinfo{journal}{\emph{Foundations and Trends{\textregistered} in Theoretical Computer Science}} \bibinfo{volume}{12}, \bibinfo{number}{1--2} (\bibinfo{year}{2016}), \bibinfo{pages}{1--274}.
\newblock


\bibitem[Wang et~al\mbox{.}(2018)]%
        {wang2018billion}
\bibfield{author}{\bibinfo{person}{Jizhe Wang}, \bibinfo{person}{Pipei Huang}, \bibinfo{person}{Huan Zhao}, \bibinfo{person}{Zhibo Zhang}, \bibinfo{person}{Binqiang Zhao}, {and} \bibinfo{person}{Dik~Lun Lee}.} \bibinfo{year}{2018}\natexlab{}.
\newblock \showarticletitle{Billion-scale commodity embedding for e-commerce recommendation in alibaba}. In \bibinfo{booktitle}{\emph{KDD}}.
\newblock


\bibitem[Wang et~al\mbox{.}(2024)]%
        {wang2023fast}
\bibfield{author}{\bibinfo{person}{Lin Wang}, \bibinfo{person}{Wenqi Fan}, \bibinfo{person}{Jiatong Li}, \bibinfo{person}{Yao Ma}, {and} \bibinfo{person}{Qing Li}.} \bibinfo{year}{2024}\natexlab{}.
\newblock \showarticletitle{Fast graph condensation with structure-based neural tangent kernel}. In \bibinfo{booktitle}{\emph{WWW}}.
\newblock


\bibitem[Wang et~al\mbox{.}(2019a)]%
        {wang2019kgat}
\bibfield{author}{\bibinfo{person}{Xiang Wang}, \bibinfo{person}{Xiangnan He}, \bibinfo{person}{Yixin Cao}, \bibinfo{person}{Meng Liu}, {and} \bibinfo{person}{Tat-Seng Chua}.} \bibinfo{year}{2019}\natexlab{a}.
\newblock \showarticletitle{Kgat: Knowledge graph attention network for recommendation}. In \bibinfo{booktitle}{\emph{KDD}}.
\newblock


\bibitem[Wang et~al\mbox{.}(2019b)]%
        {wang2019neural}
\bibfield{author}{\bibinfo{person}{Xiang Wang}, \bibinfo{person}{Xiangnan He}, \bibinfo{person}{Meng Wang}, \bibinfo{person}{Fuli Feng}, {and} \bibinfo{person}{Tat-Seng Chua}.} \bibinfo{year}{2019}\natexlab{b}.
\newblock \showarticletitle{Neural graph collaborative filtering}. In \bibinfo{booktitle}{\emph{SIGIR}}.
\newblock


\bibitem[Wang et~al\mbox{.}(2020)]%
        {wang2020disentangled}
\bibfield{author}{\bibinfo{person}{Xiang Wang}, \bibinfo{person}{Hongye Jin}, \bibinfo{person}{An Zhang}, \bibinfo{person}{Xiangnan He}, \bibinfo{person}{Tong Xu}, {and} \bibinfo{person}{Tat-Seng Chua}.} \bibinfo{year}{2020}\natexlab{}.
\newblock \showarticletitle{Disentangled graph collaborative filtering}. In \bibinfo{booktitle}{\emph{SIGIR}}.
\newblock


\bibitem[Wu et~al\mbox{.}(2019)]%
        {wu2019simplifying}
\bibfield{author}{\bibinfo{person}{Felix Wu}, \bibinfo{person}{Amauri Souza}, \bibinfo{person}{Tianyi Zhang}, \bibinfo{person}{Christopher Fifty}, \bibinfo{person}{Tao Yu}, {and} \bibinfo{person}{Kilian Weinberger}.} \bibinfo{year}{2019}\natexlab{}.
\newblock \showarticletitle{Simplifying graph convolutional networks}. In \bibinfo{booktitle}{\emph{ICML}}.
\newblock


\bibitem[Wu et~al\mbox{.}(2021)]%
        {wu2021self}
\bibfield{author}{\bibinfo{person}{Jiancan Wu}, \bibinfo{person}{Xiang Wang}, \bibinfo{person}{Fuli Feng}, \bibinfo{person}{Xiangnan He}, \bibinfo{person}{Liang Chen}, \bibinfo{person}{Jianxun Lian}, {and} \bibinfo{person}{Xing Xie}.} \bibinfo{year}{2021}\natexlab{}.
\newblock \showarticletitle{Self-supervised graph learning for recommendation}. In \bibinfo{booktitle}{\emph{SIGIR}}.
\newblock


\bibitem[Wu et~al\mbox{.}(2020)]%
        {wu2020comprehensive}
\bibfield{author}{\bibinfo{person}{Zonghan Wu}, \bibinfo{person}{Shirui Pan}, \bibinfo{person}{Fengwen Chen}, \bibinfo{person}{Guodong Long}, \bibinfo{person}{Chengqi Zhang}, {and} \bibinfo{person}{S~Yu Philip}.} \bibinfo{year}{2020}\natexlab{}.
\newblock \showarticletitle{A comprehensive survey on graph neural networks}.
\newblock \bibinfo{journal}{\emph{TNNLS}} \bibinfo{volume}{32}, \bibinfo{number}{1} (\bibinfo{year}{2020}), \bibinfo{pages}{4--24}.
\newblock


\bibitem[Xue et~al\mbox{.}(2017)]%
        {xue2017deep}
\bibfield{author}{\bibinfo{person}{Hong-Jian Xue}, \bibinfo{person}{Xinyu Dai}, \bibinfo{person}{Jianbing Zhang}, \bibinfo{person}{Shujian Huang}, {and} \bibinfo{person}{Jiajun Chen}.} \bibinfo{year}{2017}\natexlab{}.
\newblock \showarticletitle{Deep matrix factorization models for recommender systems.}. In \bibinfo{booktitle}{\emph{IJCAI}}.
\newblock


\bibitem[Xue et~al\mbox{.}(2023)]%
        {xue2023lazygnn}
\bibfield{author}{\bibinfo{person}{Rui Xue}, \bibinfo{person}{Haoyu Han}, \bibinfo{person}{MohamadAli Torkamani}, \bibinfo{person}{Jian Pei}, {and} \bibinfo{person}{Xiaorui Liu}.} \bibinfo{year}{2023}\natexlab{}.
\newblock \showarticletitle{LazyGNN: Large-Scale Graph Neural Networks via Lazy Propagation}. In \bibinfo{booktitle}{\emph{ICML}}.
\newblock


\bibitem[Ying et~al\mbox{.}(2018)]%
        {ying2018graph}
\bibfield{author}{\bibinfo{person}{Rex Ying}, \bibinfo{person}{Ruining He}, \bibinfo{person}{Kaifeng Chen}, \bibinfo{person}{Pong Eksombatchai}, \bibinfo{person}{William~L Hamilton}, {and} \bibinfo{person}{Jure Leskovec}.} \bibinfo{year}{2018}\natexlab{}.
\newblock \showarticletitle{Graph convolutional neural networks for web-scale recommender systems}. In \bibinfo{booktitle}{\emph{KDD}}.
\newblock


\bibitem[Yu et~al\mbox{.}(2022a)]%
        {yu2022graphfm}
\bibfield{author}{\bibinfo{person}{Haiyang Yu}, \bibinfo{person}{Limei Wang}, \bibinfo{person}{Bokun Wang}, \bibinfo{person}{Meng Liu}, \bibinfo{person}{Tianbao Yang}, {and} \bibinfo{person}{Shuiwang Ji}.} \bibinfo{year}{2022}\natexlab{a}.
\newblock \showarticletitle{GraphFM: Improving large-scale GNN training via feature momentum}. In \bibinfo{booktitle}{\emph{ICML}}.
\newblock


\bibitem[Yu et~al\mbox{.}(2022b)]%
        {yu2022graph}
\bibfield{author}{\bibinfo{person}{Junliang Yu}, \bibinfo{person}{Hongzhi Yin}, \bibinfo{person}{Xin Xia}, \bibinfo{person}{Tong Chen}, \bibinfo{person}{Lizhen Cui}, {and} \bibinfo{person}{Quoc Viet~Hung Nguyen}.} \bibinfo{year}{2022}\natexlab{b}.
\newblock \showarticletitle{Are graph augmentations necessary? simple graph contrastive learning for recommendation}. In \bibinfo{booktitle}{\emph{SIGIR}}.
\newblock


\bibitem[Zeng et~al\mbox{.}(2019)]%
        {zeng2019graphsaint}
\bibfield{author}{\bibinfo{person}{Hanqing Zeng}, \bibinfo{person}{Hongkuan Zhou}, \bibinfo{person}{Ajitesh Srivastava}, \bibinfo{person}{Rajgopal Kannan}, {and} \bibinfo{person}{Viktor Prasanna}.} \bibinfo{year}{2019}\natexlab{}.
\newblock \showarticletitle{GraphSAINT: Graph Sampling Based Inductive Learning Method}. In \bibinfo{booktitle}{\emph{ICLR}}.
\newblock


\bibitem[Zhang et~al\mbox{.}(2022)]%
        {zhang2022graph}
\bibfield{author}{\bibinfo{person}{Wentao Zhang}, \bibinfo{person}{Ziqi Yin}, \bibinfo{person}{Zeang Sheng}, \bibinfo{person}{Yang Li}, \bibinfo{person}{Wen Ouyang}, \bibinfo{person}{Xiaosen Li}, \bibinfo{person}{Yangyu Tao}, \bibinfo{person}{Zhi Yang}, {and} \bibinfo{person}{Bin Cui}.} \bibinfo{year}{2022}\natexlab{}.
\newblock \showarticletitle{Graph attention multi-layer perceptron}. In \bibinfo{booktitle}{\emph{KDD}}.
\newblock


\bibitem[Zhu(2005)]%
        {zhu2005semi}
\bibfield{author}{\bibinfo{person}{Xiaojin Zhu}.} \bibinfo{year}{2005}\natexlab{}.
\newblock \bibinfo{booktitle}{\emph{Semi-supervised learning with graphs}}.
\newblock \bibinfo{publisher}{Carnegie Mellon University}.
\newblock


\bibitem[Zou et~al\mbox{.}(2019)]%
        {zou2019layer}
\bibfield{author}{\bibinfo{person}{Difan Zou}, \bibinfo{person}{Ziniu Hu}, \bibinfo{person}{Yewen Wang}, \bibinfo{person}{Song Jiang}, \bibinfo{person}{Yizhou Sun}, {and} \bibinfo{person}{Quanquan Gu}.} \bibinfo{year}{2019}\natexlab{}.
\newblock \showarticletitle{Layer-dependent importance sampling for training deep and large graph convolutional networks}. In \bibinfo{booktitle}{\emph{NeurIPS}}.
\newblock


\end{thebibliography}

\clearpage
\appendix
\nobalance
\section{Appendix}

\vspace{-0.125in}
\SetAlgoNlRelativeSize{0}
\begin{algorithm}[thbp]
\SetInd{0.15em}{0.85em}
\SetKwComment{tcp}{$\triangleright$\hskip0.1em\relax}{}
\KwIn{$\text{User-item interactions } \mathcal{R} \text{; BPR batch size } B \text{; Epochs } E$} 
\KwOut{$\text{Optimized user-item embeddings } \boldsymbol{E}^{final}_{in}$}
Initialize training iteration count $k\gets 0$\;
Initialize the user-item embeddings $\boldsymbol{E}^0_{in}\sim \mathcal{N}(\mu, \sigma^2)$\;
\For{$epoch=1\dots E$}{
Compute variance reduction memory $[\boldsymbol{M}_{in}; \boldsymbol{M}_{in}']\gets[\boldsymbol{E}_{out}^{k-1},\ \frac{\partial\mathcal{L}}{\partial\boldsymbol{E}^{k-1}_{in}}], [\boldsymbol{M}_{ag};\boldsymbol{M}_{ag}']\gets\tilde{\boldsymbol{A}}[\boldsymbol{M}_{in};\boldsymbol{M}_{in}']$\tcp*[r]{$\mathcal{O}(|\mathcal{E}|d)$}
\For(\tcp*[f]{$\frac{|\mathcal{E}|}{B}$ batches}){sample $B$ interactions $\hat{\mathcal{R}}^+$ from $\mathcal{R}^+$} {
Obtain training data $\hat{\mathcal{O}}=\{(u, i, j)|(u, i)\in\hat{\mathcal{R}}^+, (u,j)\in\mathcal{R}^-\}$\;
Obtain the set of target nodes $\boldsymbol{B}=\bigcup_{(u, i, j)\in\hat{\mathcal{O}}}\{u, i, j\}$\;
Sample the random neighbors for each target node in $\boldsymbol{B}$ to obtain $\hat{\boldsymbol{A}}$\;
Obtain the forward output $\boldsymbol{E}^k_{out}$ according to Eq.~\eqref{eq:mat_vr_prop1} and Eq.~\eqref{eq:mat_vr_prop2}\tcp*[r]{$\mathcal{O}(n_BDd)$}
Compute the loss function $\mathcal{L}_{BPR}$ in Eq.~\eqref{eq:bpr_batch} and the gradients at the output layer $\frac{\partial\mathcal{L}}{\partial\boldsymbol{E}^k_{out}}$\tcp*[r]{$\mathcal{O}(Bd)$}
Compute the implicit gradients $\frac{\partial\mathcal{L}}{\partial\boldsymbol{E}^k_{in}}$ according to Eq.~\eqref{eq:back_vr_prop1} and Eq.~\eqref{eq:back_vr_prop2} \tcp*[r]{$\mathcal{O}(n_BDd)$}
Update the embedding table $\boldsymbol{E}_{in}$ with an arbitrary optimizer $\boldsymbol{E}^{k+1}_{in}\gets \text{UPDATE}(\boldsymbol{E}^{k}_{in}, \frac{\partial\mathcal{L}}{\partial\boldsymbol{E}^k_{in}})$\tcp*[r]{$\mathcal{O}(n_Bd)$}
Save $\boldsymbol{E}^k_{out}$ and $\frac{\partial\mathcal{L}}{\partial\boldsymbol{E}^k_{in}}$ to memory for the next training iteration\;
$k\gets k+1$ \tcp*[r]{The for-loop costs $\mathcal{O}(|\mathcal{E}|Dd)$}
} 
}
\textbf{return} the optimized embeddings $\boldsymbol{E}^{final}_{in} = \boldsymbol{E}^k_{in}$.
\caption{The training process of LTGNN\label{alg:ltgnn}}
\end{algorithm}
\vspace{-0.25in}

\subsection{Model Details}\label{sec:append_model} 

In this subsection, we will introduce the model details of our proposed Linear Time Graph Neural Network (LTGNN), which are not fully covered in previous Section~\ref{sec:proposed}. 

\noindent \textbf{Model Training}. The detailed training process of LTGNN is presented in Algorithm~\ref{alg:ltgnn}. 
First, we initialize the training iteration count $k$ and the trainable user-item embeddings $\boldsymbol{E}_{in}$ (line 1-2). 
Next, we repeat the outer training loop $E$ times for $E$ training epochs (lines 3-16). 
In each training epoch, we first compute the variance reduction memory for variance reduced aggregation (line 4), and then start the mini-batch training iterations (lines 5-15). 

In each training iteration, the mini-batch interactions $\hat{\mathcal{R}}^+$ are sampled from the observed user-item interactions $\mathcal{R}^{+}$. 
For every mini-batch, we first conduct negative sampling to obtain the training data $\hat{\mathcal{O}}$ (line 6), and then sample the neighbors for the target nodes in GNN aggregations (lines 7-8). 
After the negative and neighbor sampling process, we compute the forward and backward propagations of LTGNN, which are detailed in previous lines (lines 9-12). 
At the end of each training iteration, the forward and backward outputs are maintained for the incoming training iteration (line 13), enabling us to solve the PPNP fixed point with these historical computations.

\noindent \textbf{Random Aggregation with $\hat{\boldsymbol{A}}$.} For each target node in $\boldsymbol{B}$, which consists of all the users, positive items, and negative items, we sample $D$ random neighbors without repetition (including the node itself). We denote the set of random neighbors for a target node $u$ as $\hat{\mathcal{N}}(u)$, and its original neighbor set as $\mathcal{N}(u)$. We define the elements of the random adjacency matrix $\hat{\boldsymbol{A}}$ as follows:
\begin{equation}\label{eq:a_hat}
    \hat{\boldsymbol{A}}_{u, v} = \left\{ 
    \begin{aligned}
        &\frac{|\mathcal{N}(u)|}{D}\tilde{\boldsymbol{A}}_{u, v}, & \text{if $u\in\boldsymbol{B}$ and $v\in \hat{\mathcal{N}}(u)$} \\
        &0, & \text{otherwise} 
    \end{aligned}
    \right..
\end{equation}
Thus, the matrix form embedding propagation $\hat{\boldsymbol{X}} = \hat{\boldsymbol{A}}\boldsymbol{E}$ is equivalent to the node-wise random aggregation:
\begin{equation}
    \hat{\boldsymbol{x}}_u = \sum_{v\in\hat{\mathcal{N}}(u)} \frac{|\mathcal{N}(u)|}{D}\tilde{\boldsymbol{A}}_{u, v} \boldsymbol{e}_v, \quad\quad \forall u\in\boldsymbol{B}.
\end{equation}
It is clear that $\hat{\boldsymbol{x}}_u$ is an unbiased estimator of exact aggregation $\boldsymbol{x}_u = \sum_{v\in\mathcal{N}(u)} \boldsymbol{A}_{u, v}\boldsymbol{e}_v$, since 
\begin{align*}
    \mathbb{E}[\hat{\boldsymbol{x}}_u] &= \frac{|\mathcal{N}(u)|}{D}\mathbb{E}[\sum_{v\in\mathcal{N}(u)}\tilde{\boldsymbol{A}}_{u, v} \boldsymbol{e}_v\mathbb{I}(v\ |\ u)]\\
    &= \frac{|\mathcal{N}(u)|}{D}\sum_{v\in\mathcal{N}(u)}\tilde{\boldsymbol{A}}_{u, v} \boldsymbol{e}_v\mathbb{E}[\mathbb{I}(v\ |\ u)] \\ 
    &= \frac{|\mathcal{N}(u)|}{D}\sum_{v\in\mathcal{N}(u)}\tilde{\boldsymbol{A}}_{u, v} \boldsymbol{e}_v \frac{\mathcal{C}_{|\mathcal{N}(u)| - 1}^{D-1}}{\mathcal{C}_{|\mathcal{N}(u)|}^{D}} \\
    &= \sum_{v\in\mathcal{N}(u)} \boldsymbol{A}_{u, v}\boldsymbol{e}_v = \boldsymbol{x}_u,
\end{align*}
where $\mathbb{I}(v\ |\ u)$ is an indicator function that equals to $1$ when $v$ is sampled for target node $u$ and $0$ otherwise. This unbiasedness holds for both forward (line 9) and backward (line 11) computations in Algorithm~\ref{alg:ltgnn}, and is in line with previous discussions in VR-GCN~\cite{chen2018stochastic}. 

\noindent \textbf{Detailed Complexity Analysis.} In each training epoch, the efficiency bottleneck lies in the forward and backward computations in line 9 and line 11. 
According to Eq.~\eqref{eq:a_hat}, there are $n_BD$ edges in the random adjacency matrix $\hat{\boldsymbol{A}}$, so the variance-reduced neighbor aggregation has complexity $\mathcal{O}(n_BDd)$ with the help of sparse matrix multiplications. 
Thus, the overall complexity of our method is $\cO(\frac{|\cE|}{B}\cdot n_BDd) = \cO(|\cE|Dd)$, since the inner training loop in lines 5-15 repeats $\frac{|\cE|}{B}$ times, and $B$ and $n_B$ have the same order. 
Given that $D$ is a small constant, the complexity becomes linear to the number of edges $|\cE|$, demonstrating high scalability potential.
 
It is also noteworthy that our variance-reduced aggregation strategy does not affect the linear complexity of LTGNN. 
Particularly, the variance-reduced aggregations in lines 9 and 11 have the same complexity as vanilla neighbor sampling methods in PinSAGE, while the extra computational overhead of computing the variance reduction memory is $\mathcal{O}(|\mathcal{E}|d)$, which is as efficient as the simplest MF model (line 2). 
This means our proposed variance reduction approach reduces the random error without sacrificing the training efficiency. 

\noindent \textbf{Model Inference.} After the model training process described in Algorithm~\ref{alg:ltgnn}, we have several options to predict user preference with the optimized input embeddings $\boldsymbol{E}^{final}_{in}$. 
The simplest solution is to directly infer the output embeddings with Eqs.~\eqref{eq:mat_vr_prop1}~-~\eqref{eq:mat_vr_prop2}, and then predict the future user-item interaction with an inner product. However, the training process may introduce small errors due to reusing the historical computations, which could cause $\boldsymbol{E}^{final}_{out}$ to deviate from the exact PPNP fixed-point $\boldsymbol{E}^{final}_{PPNP}$ w.r.t. $\boldsymbol{E}^{final}_{in}$. 
Therefore, a possible improvement is to compute $\boldsymbol{E}^{final}_{PPNP}$ with APPNP layers in Eqs.~\eqref{eq:mp_in}~-~\eqref{eq:mp_iter}. 
This strategic choice can reduce the error due to reusing the previous computations, despite slightly compromising the behavior consistency between training and inference.

We find that both inference choices can accurately predict the user preference, and choosing either of them does not have a significant impact on the recommendation performance. 
In practice, we compute $\boldsymbol{E}^{final}_{PPNP}$ with a 3-layer APPNP which takes $\boldsymbol{E}^{final}_{in}$ as the input.

\subsection{Parameter and Evaluation Settings}\label{sec:append_setting}
We implement the proposed LTGNN using PyTorch
and PyG 
libraries. 
We strictly follow the settings of NGCF~\citep{wang2019neural} and LightGCN~\citep{he2020lightgcn} to implement our method and the scalable LightGCN variants for a fair comparison. 
All the methods use an embedding size of 64, a BPR batch size of 2048, and a negative sample size of 1.
For the proposed LTGNN, we tune the learning rate from $\{\text{5e-4}, \text{1e-3}, \text{1.5e-3}, \text{2e-3}\}$ and the weight decay from $\{\text{1e-4}, \text{2e-4}, \text{3e-4}\}$. 
We employ an Adam~\cite{kingma2014adam} optimizer to minimize the objective function. 
For the coefficient $\alpha$ in PPNP, we perform a grid search over the hyperparameter in the range of [0.3, 0.7] with a step size of 0.05. 
To ensure the scalability of our model, the number of propagation layers $L$ is fixed to 1 by default, and the number of sampled neighbors $D$ is searched in $\{5, 10, 15, 20\}$. 
For the GNN-based baselines, we follow their official implementations and suggested settings in their papers. 
For the LightGCN variants with scalability techniques, the number of layers $L$ is set based on the best choice of LightGCN, and we search other hyperparameters in the same range as LTGNN and report the best results.

In this paper, we adopt two widely used evaluation metrics in recommendations: Recall@K and Normalized Discounted
Cumulative Gain (NDCG@K)~\citep{wang2019neural, he2020lightgcn}. 
We set the K=20 by default, and we report the average result for all test users. 
All the experiments are repeated five times with different random seeds, and we report the average results. 

\subsection{Additional Experiments}\label{sec:append_exp}
\noindent \textbf{Comparison to GTN.} Despite numerous works being orthogonal to our contributions and not tackling the scalability issue, as highlighted in Section~\ref{sec:exp_setting}, we remain receptive to comparing our method with more recent, accuracy-focused baselines in the field. 
We compare our proposed LTGNN with GTN~\cite{fan2022graph}, one of the latest GNN backbones for recommendations, and the results are presented in Table.~\ref{tab:gtn_acc} and Table.~\ref{tab:gtn_eff}.

\begin{table}[htb]
\caption{Recommendation performance comparison results with extra baseline GTN. Results marked with (*) are obtained in GTN's official settings.}
\vskip -0.15in
\label{tab:gtn_acc}
\begin{center}
\scalebox{0.65}
{
\begin{tabular}{l|c c|c c|cc}
\hline
\textbf{Dataset}  & \multicolumn{2}{c|}{\textbf{Yelp2018}} & \multicolumn{2}{c|}{\textbf{Alibaba-iFashion}} & \multicolumn{2}{c}{\textbf{Amazon-Large}} \\ \hline
\textbf{Method}  & \textbf{Recall@20} & \textbf{NDCG@20} & \textbf{Recall@20} & \textbf{NDCG@20} & \textbf{Recall@20} & \textbf{NDCG@20}  \\ \hline\hline
GTN & 0.0679* & 0.0554* & 0.0994* & 0.0474* & OOM & OOM \\ 
LTGNN & \textbf{0.0681}* & \textbf{0.0562}* & \textbf{0.1079}* & \textbf{0.0510}* & \textbf{0.0294} & \textbf{0.0259} \\
\hline\hline
\end{tabular}
}
\end{center}
\vskip -0.15in
\end{table}

\begin{table}[htb]
\caption{Running time comparison results with extra baseline GTN. Results marked with (*) are obtained in GTN's official settings.}
\vskip -0.15in
\label{tab:gtn_eff}
\begin{center}
\scalebox{0.85}
{
\begin{tabular}{l|c|c|c}
\hline
\textbf{Dataset}  & \multicolumn{1}{c|}{\textbf{Yelp2018}} & \multicolumn{1}{c|}{\textbf{Alibaba-iFashion}} & \multicolumn{1}{c}{\textbf{Amazon-Large}} \\ \hline
\textbf{Method}  & \textbf{Running Time} & \textbf{Running Time} & \textbf{Running Time} \\ \hline\hline
GTN & 97.8s* & 99.6s* & OOM \\
LTGNN & \textbf{24.5s}*	& \textbf{22.4s}* & \textbf{705.9s} \\
\hline\hline
\end{tabular}
}
\end{center}
\end{table}

To ensure a fair comparison, we closely followed the official settings of GTN~\footnote{~\url{https://github.com/wenqifan03/GTN-SIGIR2022}}, which differ from our evaluation settings detailed in Section~\ref{sec:exp_setting} in embedding size and BPR batch size. 
From the experiment result, we find that our proposed LTGNN shows advantages in both recommendation performance and scalability, even in comparison with one of the latest accuracy-driven GNN backbones for recommendations.

\begin{figure}[ht]
\vskip -0.1in
\centering
\subfigure{
\begin{minipage}[t]{0.43\linewidth}
\centering
\includegraphics[width=\linewidth]{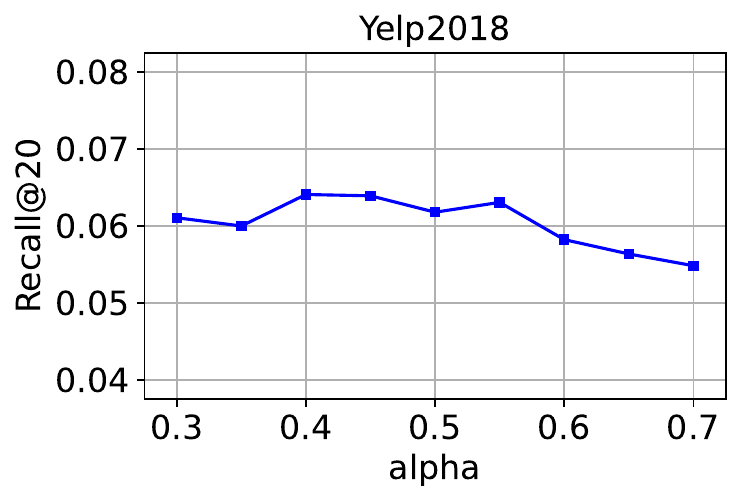}
\end{minipage}
\begin{minipage}[t]{0.43\linewidth}
\centering
\includegraphics[width=\linewidth]{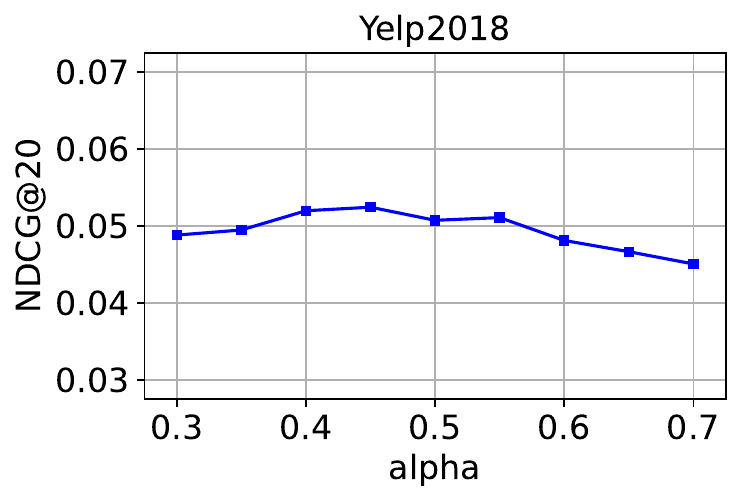}
\end{minipage}
}

\subfigure{
\begin{minipage}[t]{0.43\linewidth}
\centering
\includegraphics[width=\linewidth]{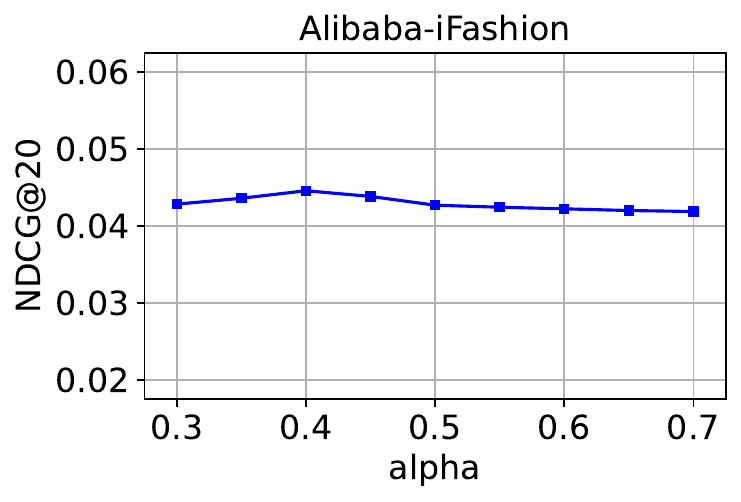}
\end{minipage}
\begin{minipage}[t]{0.43\linewidth}
\centering
\includegraphics[width=\linewidth]{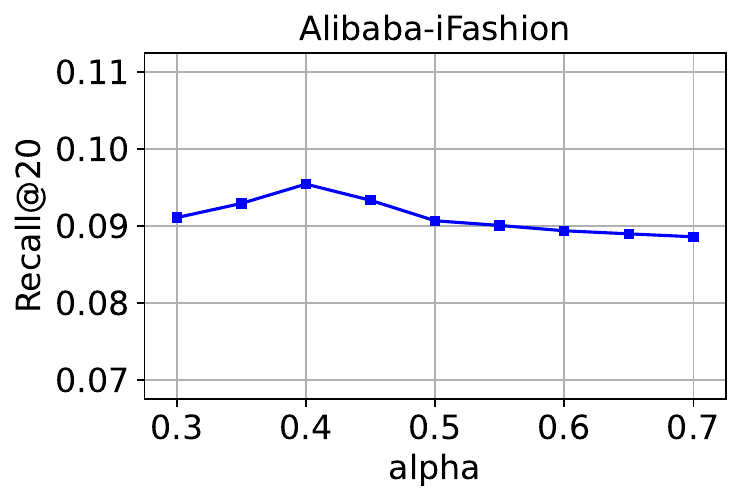}
\end{minipage}
}
\vskip -0.2in
\caption{The effect of hyper-parameter $\alpha$ under Recall@20 and NDCG@20 metrics.}
\label{fig:perf_appnp_alpha}
\vskip -0.1in
\end{figure}

\noindent \textbf{Effect of Hyperparameter $\alpha$.} Similar to APPNP~\cite{gasteiger2018predict}, the teleport factor $\alpha$ in LTGNN controls the trade-off between capturing long-range dependencies and graph localities, directly impacting recommendation performance. 
As shown in Fig.~\ref{fig:perf_appnp_alpha}, we find that LTGNN is rather robust to $\alpha$, as the performance does not drop sharply in $\alpha\in [0.35, 0.55]$. 
In practice, we can set $\alpha=0.5$, which is the midpoint of capturing long-range dependency and preserving the local structure expressiveness, and conduct a rough grid search to find the optimal setting of $\alpha$.

\noindent \textbf{Effect of Different Variance Reduction Designs}. We developed an efficient variance reduction (EVR) mechanism in Section~\ref{sec:vr}, which can be applied to both forward and backward computations. 
To evaluate the impact of different variance reduction designs on the recommendation performance, we compare four variants of LTGNN, namely \emph{LTGNN-NS}, \emph{LTGNN-FVR}, \emph{LTGNN-BVR}, and \emph{LTGNN-BiVR}. 
LTGNN-NS is the baseline model that uses implicit modeling and vanilla neighbor sampling, while LTGNN-FVR, LTGNN-BVR, and LTGNN-BiVR use forward variance reduction, backward variance reduction, and bidirectional variance reduction, respectively. 

\begin{table}[htb]
\vskip -0.1in
\caption{Ablation study on different variance reduction designs.}
\vskip -0.15in
\label{tab:vr_design}
\begin{center}
\scalebox{0.85}
{
\begin{tabular}{l|c c|c c}
\hline
\textbf{Dataset}  & \multicolumn{2}{c|}{\textbf{Yelp2018}} & \multicolumn{2}{c}{\textbf{Alibaba-iFashion}} \\ \hline
\textbf{Method}  & \textbf{Recall@20} & \textbf{NDCG@20} & \textbf{Recall@20} & \textbf{NDCG@20}  \\ \hline\hline
LTGNN-NS & 0.06285 & 0.05137 & 0.08804 & 0.04147 \\
LTGNN-BVR & 0.06309& 0.05160 & 0.08878 & 0.04185 \\
LTGNN-BiVR & \underline{0.06321} & \underline{0.05194} & \underline{0.09241} & \underline{0.04338} \\
LTGNN-FVR & \textbf{0.6380} & \textbf{0.05224} & \textbf{0.09335} & \textbf{0.04387} \\
\hline\hline
\end{tabular}
}
\end{center}
\vskip -0.15in
\end{table}

As can be seen from Table.~\ref{tab:vr_design}, LTGNN variants with variance reduction outperform LTGNN-NS without variance reduction, demonstrating the effectiveness of our proposed EVR approach. Moreover, we observe that forward variance reduction (FVR) alone is sufficient to achieve satisfactory recommendation performance. Surprisingly, bidirectional variance reduction does not outperform forward variance reduction. We intend to investigate the reason behind this phenomenon and explore the potential of bidirectional variance reduction (BiVR) in our future work.

\noindent \textbf{Detailed Efficiency Analysis}. As discussed in Section~\ref{sec:vr}, LTGNN has linear time complexity with respect to the number of edges $|\mathcal{E}|$ in the user-item interaction graph, which is comparable to MF. 
However, as shown in Table.~\ref{tab:comparsion_scale}, the running time of LTGNN is worse than MF on the large-scale Amazon-Large dataset, which seems to contradict our complexity analysis. 
To understand this phenomenon and demonstrate LTGNN’s scalability advantage, we perform a comprehensive efficiency analysis of LTGNN and MF, which identifies the main source of overhead and suggests potential enhancements.

From the results presented in Table.~\ref{tab:detailed_eff}, we have the following observations:
\begin{itemize}
    \item For both MF and LTGNN, the computation cost of negative sampling is negligible, which has a minimal impact on the total running time.
    \item  For LTGNN, excluding the neighbor sampling time, the model training time is linear in the number of edges $|\mathcal{E}|$, which is similar to MF. This indicates LTGNN’s high scalability in large-scale and industrial recommendation scenarios.
    \item Besides, the computational overhead incurred by maintaining and updating the memory spaces for variance reduction is insignificant, which validates the high efficiency of our proposed EVR approach.
    \item For LTGNN, the main source of extra computational overhead is the neighbor sampling process for the target nodes, which accounts for more than 50\% of the total running time, preventing LTGNN from achieving a perfect linear scaling behavior. This is an implementation issue that can be improved by better engineering solutions.
\end{itemize}

To further enhance the efficiency of LTGNN, we can adopt some common engineering techniques. For example, we can follow the importance sampling approach in PinSAGE~\cite{ying2018graph}, which assigns a fixed neighborhood for each node in the interaction graph, avoiding the expensive random sampling process. We plan to leave this as a future work. 

\begin{table}[!b]
\vskip -0.1in
\caption{Detailed efficiency comparison of LTGNN with MF.}
\vskip -0.15in
\label{tab:detailed_eff}
\begin{center}
\scalebox{0.78}
{
\begin{tabular}{lc|c|c|c}
\hline
\multirow{2}{*}{\textbf{Model}} & \multirow{2}{*}{\textbf{Stage}} & \textbf{Yelp2018} & \textbf{Alibaba-iFashion} & \textbf{Amazon-Large} \\ 
 & & $|\mathcal{E}|=1.56m$ & $|\mathcal{E}|=1.61m$ & $|\mathcal{E}|=15.24m$ \\
 \hline\hline
\multirow{3}{*}{MF} & Neg. Samp. & 0.12s & 0.21s & 1.23s \\
                                & Training & \underline{4.19s} & \underline{4.39s} & \underline{126.01s} \\
                                & \textbf{Total} & 4.31s & 4.60s & 127.24s \\
\hline
\multirow{5}{*}{LTGNN} & Neg. Samp. & 0.12s & 0.21s & 1.23s \\
                                & Neigh. Samp. & 7.98s & 6.48s & 512.55s \\
                                & Training & \underline{6.62s} & \underline{6.99s} & \underline{192.13s} \\
                                & Memory Access & <0.005s & <0.005s & <0.005s \\
                                & \textbf{Total} & 14.72s & 13.68s & 705.91s \\
\hline\hline
\end{tabular}
}
\end{center}
\vskip -0.15in
\end{table}

\end{document}